\documentclass[useAMS,usenatbib]{mn2e}
\usepackage{graphics, graphicx, amsmath, amssymb}
\usepackage{tabularx}
\usepackage{verbatim}

\title[Chondrule Transport]{Chondrule Transport in Protoplanetary Disks}
\author[A. Z. Goldberg, J. E. Owen \& E. Jacquet]{Aaron Z. Goldberg$^{1,2}$\thanks{E-mail: goldberg@cita.utoronto.ca}, James E. Owen$^{1,3}$\thanks{Hubble Fellow} and Emmanuel Jacquet$^{1,4}$\\
	$^{1}$Canadian Institute for Theoretical Astrophysics, 60 St. George Street, Toronto, M5S 3H8, Canada.\\
		$^{2}$Department of Physics and Astronomy, McMaster University, 1280 Main Street W., Hamilton, Ontario, Canada, L8S 4M1.\\
	$^{3}$Institute for Advanced Study, Einstein Drive, Princeton NJ, 08540, USA\\
	$^{4}$Institut de Min\'{e}ralogie, de Physique des Mat\'{e}riaux et de Cosmochimie, Mus\'{e}um National d'Histoire Naturelle, \\CP52, 57 rue Buffon, 75005 Paris, France.}

\begin{document}
	\include{journals_mnras}
	\pagerange{\pageref{firstpage}--\pageref{lastpage}} \pubyear{2014}
	
	\maketitle
	
	\label{firstpage}
	
	\begin{abstract}
	Chondrule formation remains one of the most elusive early Solar System events. Here, we take the novel approach of employing numerical simulations to investigate chondrule origin beyond purely cosmochemical methods. We model the transport of generically-produced chondrules and dust in a 1D viscous protoplanetary disk model, in order to constrain the chondrule formation events. For a single formation event we are able to match analytical predictions of the memory chondrule and dust populations retain of each other (complementarity), finding that a large mass accretion rate ($\gtrsim 10^{-7}$~M$_\odot$~yr$^{-1}$) allows for delays on the order of the disk's viscous timescale between chondrule formation and chondrite accretion. Further, we find older disks to be severely diminished of chondrules, with accretion rates $\lesssim 10^{-9}$~M$_\odot$~yr$^{-1}$ for nominal parameters. We then characterize the distribution of chondrule origins in both space and time, as functions of disk parameters and chondrule formation rates, in runs with continuous chondrule formation and both static and evolving disks. Our data suggest that these can account for the observed diversity between distinct chondrite classes, if some diversity in accretion time is allowed for. 
	\end{abstract}
	
	
	\section{Introduction}
	The earliest witnesses of the physical and chemical processes that took place during the formation of the Solar System, 4.57 billion years ago, belong to the type of meteorites called chondrites. Chondrites are primitive, having mostly escaped differentiation processes on their parent asteroids, and are composed of various roughly millimeter-sized inclusions native to the solar protoplanetary disk set in a fine-grained matrix. While the oldest among these inclusions are the \textit{refractory inclusions}, presumably high-temperature condensates from the earliest phases of the disk, the most abundant are millimeter-sized silicate spheroids known as \textit{chondrules}, which formed as a result of the solidification of molten droplets \citep{Amelinetal2010, ConnollyDesch2004}. The matrix itself is a composite mixture of micron-size presolar components, protoplanetary disk condensates, and possible by-products of chondrule-forming events \citep{Hussetal2005}.

	Although theories abound, the nature of the chondrule formation events (CFEs) remains very mysterious \citep[e.g.][]{Bischoff1998, ConnollyDesch2004}.
	There are two main categories for CFE models, more or less equally backed in the latest literature: ``planetary'' and ``nebular''. The former may require phenomena such as collisions between planetesimals in the disk \citep{Symesetal1998,Asphaugetal2011,SandersScott2012,Johnsonetal2015}, and a popular example of the latter relies upon shock waves passing through the disk \citep{DeschConnolly2002,MorrisDesch2010,Morrisetal2012,Boleyetal2013}.
	
	A key constraint is that chondrules, even in single chondrites, have ages spread over 0-3 Myr after the condensation of refractory inclusions.  Not only does this suggest that CFEs occurred repetitively over the evolution timescale of the solar protoplanetary disk, as disks typically last $\sim$ 3 Myr around young stars \citep{WilliamsCieza2011}, this also indicates that chondrules did not generally accrete immediately after formation, but spent a few Myr as free-floating objects in the gaseous disk. Chondrules may have been significantly redistributed by aerodynamic forces \citep[e.g.][]{CuzziWeidenschilling2006,Jacquet2014review}. At the same time, chondrules seem to form distinct populations in different chondrite groups, constraining any disk-wide mixing \citep{Jones2012,AlexanderEbel2012}.
	
	Transport of chondrules may also be crucial to understand the compositional diversity of chondrites \citep{Jacquet2014review}. Indeed chondrites are not a uniform class, but are currently comprised of 14 chemically, isotopically, and petrographically distinct groups, which to first order may be subsumed in two superclans \citep{Kallemeynetal1996,Warren2011b}: \textit{carbonaceous chondrites} (CCs; matrix-rich, with mostly solar composition  -- especially for the CI group -- e.g. in terms of the Mg/Si ratio),	and non-carbonaceous chondrites (including ordinary and enstatite chondrites, which are matrix-poor and further from solar). One school of thought is that the chondrite spectrum reflects varying mixing proportions between a high-temperature component dominated by chondrules and a low-temperature component believed to be CI composition matrix \citep{Anders1964,Zandaetal2006,Zandaetal2012}; this is the \textit{two-component model}. Yet this view is challenged by evidence that matrix in carbonaceous chondrites is non-solar, e.g. in terms of its Mg/Si and Fe/Si ratios, and neither are the coexisting chondrules, although the bulk carbonaceous chondrites \textit{are} close to solar for these elemental ratios \citep{Blandetal2005,HezelPalme2010}. This matrix-chondrule \textit{complementarity} -- if not due to analytical artifacts or parent body processes \citep{Zandaetal2012} -- suggests a genetic relationship between the two, with chemical exchange in a closed-system reservoir \citep{HezelPalme2010,Palmeetal2015}. For example, if Si evaporated from the chondrules, it could have preferentially recondensed on matrix grains owing to their larger total surface area. This certainly constrains transport \textbf{of solids} since chondrule formation \citep{Jacquetetal2012S}.

	Clearly, the transport of chondrules is important to understand, not only to better comprehend chondrites as ``end products'' of disk processes, but as a constraint for CFEs themselves. Yet, despite considerable amount of work devoted to the transport of solid particles in general, as part of the effort to understand chondrite accretion \citep[see e.g. reviews by][]{CuzziWeidenschilling2006,ChiangYoudin2010}, the transport of chondrules \textit{per se} has been little investigated specifically in the literature, save as part of the -- purely analytical -- synthesis on chondrite component transport by \citet{Jacquetetal2012S}. In contrast, several numerical studies on the transport of refractory inclusions already exist \citep[e.g.][]{Cuzzietal2003,Hu2010,Ciesla2010,Bossetal2012,YangCiesla2012}. This is certainly understandable: the origin of chondrules is so ill-understood (even compared to refractory inclusions) and the disk properties similarly uncertain that such an endeavour may seem premature or arbitrary at best. Yet, as time goes by, it does appear that chondrule formation and disk transport stand little chance to be solved entirely separately, and that a first attempt to mutually constrain these interwoven problems in numerical simulations has to be made. This is the purpose of this work.
	
	In this work, we simulate the transport of chondrules, chondrule precursors, and dust grains in $1D$ models of ``conventional" turbulent gaseous disks and monitor the chondrite composition expected at any given time and heliocentric distance to compare with observations. We first assume a fixed, initial pair of chondrule and dust populations, focusing on the question of matrix-chondrule complementarity. We then allow for continuous chondrule formation, following simple prescriptions throughout the simulation time, to understand the diversity of origins of components (age, heliocentric distance of formation) present in individual chondrites.

	The outline of this paper is as follows. In Section \ref{section-1D-model}, we outline the relevant equations for gas and solid dynamics in the protoplanetary disk, along with an explanation of our numerical methods. Section \ref{Dust top hat model} contains our model of investigating the evolution of a single chondrule and dust population following a single CFE, and Section \ref{section-continuous-CFE} extends these calculations to disks with multiple CFEs. We discuss the implications of our results in Section \ref{section-discussions}, and, in Section \ref{section-conclusions}, we conclude.

	\section{Disk model and methods}
	\label{section-1D-model}
	We consider an axisymmetric disk in cylindrical coordinates, with $R$ the heliocentric distance and $z$ the height above the midplane. We work within the thin-disk formalism appropriate for protoplanetary disks \citep[e.g.][]{Pringle1981} and work with vertically averaged quantities.  The  surface density of gas, dust, or chondrules is defined by
	\begin{equation}
	\Sigma(R)\equiv\int_{-\infty}^{+\infty}\rho\mathop{\mathrm{d}z},
	\label{gas surface density}
	\end{equation} where $\rho(R,z)$ is the mass density.

	\subsection{Gas disk}
	
	At radii $R\gtrsim 0.5$~AU, the disk is passively heated \citep[e.g.][]{ChiangGoldreich1997,DAlessio2001} and as such the temperature ($T$) profile is taken to be a time-independent power law of heliocentric distance $T\propto R^{-q}$, with $q=0.5$ \citep{KenyonHartmann1987} unless otherwise noted. For a vertically isothermal disk the scale height ($H$) is given by $H=c_s/\Omega$, where $c_s
	$ is the isothermal sound speed and $\Omega\mathbf{=\sqrt{GM_\odot/R^3}}$ is the Keplerian angular velocity. Following \citet{Owen2014} we set the \textbf{aspect ratio} 
	in all \textbf{of} our calculations to: 
	\begin{equation}
	\frac{H}{R}=0.04\left(\frac{R}{1~{\rm AU}}\right)^{\mathbf{1/2-q/2}}.
	\end{equation}  
	
	Owing to turbulent angular momentum transport, the gas surface density evolves following \citep{Pringle1981}
	\begin{equation}
	\frac{\partial \Sigma_\mathrm{g}}{\partial t}=\frac{3}{R}\frac{\partial}{\partial R}\left[R^{1/2}\frac{\partial}{\partial R}\left(R^{1/2}\nu\Sigma_\mathrm{g})\right)\right],
	\label{1D gas evolution equation}
	\end{equation}
	where the ``turbulent viscosity" is given by the $\alpha$ formalism \citep{ShakuraSunyaev1973}, such that: 
	\begin{equation}
	\nu(R)\equiv\alpha c_s H .
	\end{equation}
	For constant $\alpha$ we find $\nu\propto R$, which is consistent with the observational diagnostics of disk evolution \citep[e.g.][]{Hartmannetal1998,Andrews2009}. Following \citet{Owenetal2011} we set $\alpha=2.5\times10^{-3}$ to match the disk lifetimes and accretion rates within the X-ray photoevaporation model. We note here that, while we have chosen particular scales for many of the models we discuss here (steady-disk models), our results can be re-scaled to other chosen values of $\alpha$ and $H/R$. As such we choose to work with natural scales of the disk.  
	
	Specifically, a natural timescale of the problem is the viscous timescale $t_{\rm vis}(R)$ (namely the time over which a gas parcel at distance $R$ is accreted by the Sun), given by:
	\begin{eqnarray}
	t_{\mathrm{vis}}(R)&=& \frac{R^2}{\nu} \nonumber \\
	&=&0.04 \mathrm{Myr} \left(\frac{R}{1\:\rm AU}\right)^{1/2+q}\left(\frac{\mathbf{(H/R)}(1\: \rm AU)}{0.04}\right)^{-2}\nonumber\\
	&&\left(\frac{\alpha}{2.5\times10^{-3}}\right)^{-1} .
	\end{eqnarray}
	
	The strength of the turbulence (and as such the value of $\alpha$) is very uncertain, even at the order-of-magnitude level. Therefore, most of the times discussed in the results will be normalized to $t_{\rm vis}$ so as to be independent of $\alpha$ (for a given accretion rate).
	
	For timescales longer than $t_{\rm vis}(R)$, the surface density should approximate the steady solution with a uniform mass accretion rate $\dot{M}\equiv-2\pi R\Sigma_\mathrm{g} u_R$ ($u_R$ being the gas radial velocity), given by:
	\begin{equation}
	\Sigma_\mathrm{g}=\frac{\dot{M}}{3\pi\nu}\left[1-\left(\frac{R_{*}}{R}\right)^{1/2}\right],
	\label{steady disk}
	\end{equation}
	where $R_*$ is the disk inner edge, and a zero-torque boundary condition is applied.  Therefore, with our setup at radii $\gg R_*$ the surface density follows $\Sigma\propto R^{-1}$.
	
	\subsection{Evolution of solids}
	
	A population of solids, whether chondrules, chondrule precursors, or dust grains, evolves in the disk due to (i) advection by the mean gas flow, (ii) turbulent diffusion, and (iii) drift because of finite size leading to partial decoupling with the gas.  The evolution of the surface density $\Sigma_s$ of any solid population obeys \citep[e.g.][]{Jacquetetal2012S}:
	\begin{equation}
	\frac{\partial \Sigma_\mathrm{s}}{\partial t}=\frac{1}{R}\frac{\partial}{\partial R}\left\{R\left[D\Sigma_\mathrm{g}\frac{\partial}{\partial R}\left(\frac{\Sigma_\mathrm{s}}{\Sigma_\mathrm{g}}\right)-\Sigma_\mathrm{s}\left(u_R+v_{\mathrm{drift}}\right)\right]\right\},
	\label{1D dust evolution equation}
	\end{equation}
	where $D=\nu/Sc$ is the radial gas diffusion coefficient, $Sc$ the radial Schmidt number parametrizing the strength of the dust diffusion, henceforth taken to be 1(unless otherwise noted), and
	\begin{equation}
	v_{\mathrm{drift}}=\frac{\tau}{\rho}\frac{\partial P}{\partial R}
	\end{equation}
	is the solid particle radial drift velocity arising from the gas pressure gradient. $P(R)=\rho c_\mathrm{s}^2
	$ is the midplane gas pressure and
	\begin{equation}
	\tau=\sqrt{\frac{\pi}{8}}\frac{\rho_\mathrm{s} a}{\rho c_\mathrm{s}}
	\end{equation}
	is the stopping time (due to gas drag) of a solid particle, with $\rho_\mathrm{s}$ and $a$ the particle's internal density and radius, respectively \citep{Jacquetetal2012S}. Throughout this paper, we take $\rho_\mathrm{s} a=0.1\,\mathrm{g}/\mathrm{cm}^2$ for (millimeter-size) chondrules and aggregates, and $\rho_\mathrm{s} a=1\times10^{-4}\,\mathrm{g}/\mathrm{cm}^2$ for (micron-size) dust grains. Note that we work with solid/gas ratios $\ll 1$ so that our results are linear functions of the initial solid abundances, and thus relative variations of solid abundances are independent of any assumption about their initial abundances.

	We can then define the dimensionless gas-solid decoupling parameter 
	\begin{equation}
	S\equiv \frac{\Omega\tau}{\alpha}=\frac{\pi}{2}\frac{\rho_\mathrm{s} a}{\Sigma \alpha}
	\end{equation}
	as a measure of importance of the drift contribution \citep{Jacquetetal2012S}. For a steady disk, using Equation (\ref{steady disk}), one finds that at $R\gg R_*$:
	\begin{eqnarray}
	\label{S steady}
	S &=& \frac{3\pi^2}{2}\frac{\rho_\mathrm{s} ac_s^2}{\dot{M}\Omega}\nonumber\\
	&\approx& 0.2 \left(\frac{\rho_sa}{0.1\:\rm g~cm^{-2}}\right)\left(\frac{\dot{M}}{10^{-8}\:\rm M_\odot/yr}\right)^{-1}\nonumber\\&\times&\left(\frac{\mathbf{(H/R)}(1\:\rm AU)}{0.04}\right)^2\left(\frac{R}{1\:\rm AU}\right)^{3/2-q},
	\end{eqnarray} 
	which is independent of $\alpha$ for a given value of $\dot{M}$. 
	
	No sink term corresponding to chondrite accretion is considered here. Nevertheless, at each time and radial location, one can define a \textit{potential chondrite} composition that would result from partial accretion of local material. We assume that the potential chondrite's chondrule/matrix makeup is representative of the location in question; that is, we ignore any possible ``accretion bias" as argued by \citet{Jacquetetal2012S,Jacquet2014size}. This will give us insight into how chondrite composition evolves in time and space, and whether there is a spatiotemporal ``window" matching the observations.

	\subsection{Quantifying matrix-chondrule complementarity}
	
	\label{section-complementarity}
	
	As part of the evaluation of the ``potential chondrite'' composition, we will seek to quantify the possible complementarity between chondrules and matrix as mentioned in the introduction. Complementarity cannot simply be a function of how close the bulk composition is to solar (that is, close to the CI chondrites, which are deemed to best represent solar abundances), for the individual chondrules and matrix may also be so close to solar that this is not a real constraint on their genetic relationships \citep[not to mention possible analytical biases or secondary effects;][]{Zandaetal2012}. Thus, we need to factor out the original closeness of the different components to solar. Hence, we shall define here, for a fiducial chemical element $X$, a complementarity parameter $\zeta$ as:
	\begin{equation}
	\zeta=\frac{[X]_{\mathrm{ch}}-[X]_{\mathrm{bulk}}}{[X]_{\mathrm{mx}}-[X]_{\mathrm{bulk}}}\times\frac{[X]_{\mathrm{mx}}-[X]_{\rm CI}}{[X]_{\mathrm{ch}}-[X]_{\rm CI}},
	\label{zeta equation}
	\end{equation}
	where $[X]_{\mathrm{ch}}$, $[X]_{\mathrm{mx}}$, and $[X]_{\mathrm{bulk}}$  are the abundances (by mass or normalized to some major element)
	of element $X$ in the chondrules, matrix, and whole-rock of a sample, respectively, with $[X]_{\rm CI}$ being the CI value. The motivation behind this specific form is that, assuming that there is a value $r_{\rm CI}$ of the matrix/chondrule ratio $r$ for which a mix of the observed chondrules and matrix is solar, $\zeta$ reduces to $r_{\rm sample}/r_{\rm CI}$, which is directly measurable in simulations. It further does not depend on the chosen initial chondrule/dust ratio, nor any chemical assumption in the simplest case of a single pair of chondrule and dust populations as investigated in Section \ref{Dust top hat model}. $\zeta$ is of course unity if the bulk chondrite is solar in $X$. As such, the closer $\zeta$ is to $1$, the more complementary a sample is. 
	
	In order to get a feel for real values of $\zeta$, we take the Mg/Si ratio as our $[X]$ (but the calculations herein are not \textbf{unique} to this specific chemical parameter and thus should apply to any claimed complementary relationship). Compositional data for a number of chondrite classes are tabulated in Table \ref{Complementarity table}. We see that CCs tend to have $\zeta$ within $\sim10\%$ of $1$, with the CV chondrites deviating most from this ($\zeta=0.7$). In contrast, enstatite and ordinary chondrites have negative values because both their chondrules and matrices have a subsolar Mg/Si ratio (so that no combination thereof can restore a CI chondritic composition); again complementarity is more a property of carbonaceous chondrites than of chondrites in general. In this paper, unless otherwise noted, we will thus adopt the convention that a chondrite is ``complementary" if its $\zeta$ lies between 0.7 and 1.3.
	
	\begin{table}
		\label{Complementarity table}
		\caption{Magnesium to silicon ratios for samples of the carbonaceous chondrite types CR, CV, CO, and CM, and three non-carbonaceous chondrites. All values are normalized to CI elemental abundances. Mg/Si is reported for chondrules, matrix, and whole-rock values for each chondrite, and used to calculate the complementarity parameter $\zeta$. Unless the reference mentioned a specific CI Mg/Si value, the CI value chosen for normalization (0.90) was taken from  \citet{PalmeJones2005}. \textsuperscript{1}\citet{HezelPalme2010}, \textsuperscript{2}\citet{Klerneretal2001}, \textsuperscript{3}\citet{Ebeletal2008}, \textsuperscript{4}\citet{LoddersFegley1993}, \textsuperscript{5}\citet{McSweenRichardson1977}, \textsuperscript{6}\citet{Zolenskyetal1993}, \textsuperscript{7}\citet{Mason1963}, \textsuperscript{8}\citet{RubinWasson1987}, \textsuperscript{9}\citet{GrossmanBrearley2005}, \textsuperscript{10}\citet{Ahrensetal1973}, \textsuperscript{11}\citet{Hussetal2005}, \textsuperscript{12}\citet{Grossmanetal1985}, \textsuperscript{13}\citet{Rambaldietal1984}, \textsuperscript{14}\citet{ElGoresyetal1988}, \textsuperscript{15}\citet{Berlin2009}, \textsuperscript{16}\citet{GrossmanWasson1983}.}
		\centering
		\begin{center}
			\begin{tabularx}{\linewidth}{|>{\centering\hsize=1.1\hsize}X|>{\centering\hsize=1.2\hsize}X|>{\centering\hsize=1.0\hsize}X|>{\centering\hsize=1.0\hsize}X|>{\centering\hsize=1.0\hsize}X|>{\centering\arraybackslash\hsize=0.7\hsize}X|}
				\hline
				Type & Name & $\mathrm{Mg}/\mathrm{Si}_{\mathrm{CI}}$ (chondrule) & $\mathrm{Mg}/\mathrm{Si}_{\mathrm{CI}}$ (matrix) & $\mathrm{Mg}/\mathrm{Si}_{\mathrm{CI}}$ (whole-rock) & $\zeta$ \\ \hline
				CR & Renazzo & 1.15\textsuperscript{1,2,3} & 0.71\textsuperscript{2,4,5,6} & 1.01\textsuperscript{7} & 0.93 
				\\ \hline
				CV & Vigarano & 1.18\textsuperscript{8} & 0.87\textsuperscript{2,5,6} & 1.03\textsuperscript{7} & 0.70 \\ \hline
				CO & Kainsaz & 1.19\textsuperscript{1} & 0.85\textsuperscript{1,5,9} & 0.99\textsuperscript{10} & 1.07 \\ \hline
				CM & El-Quss Abu Said & 1.23\textsuperscript{1} & 0.73\textsuperscript{1} & 1.00\textsuperscript{4,11} & 0.99 \\ \hline
				E & Qingzhen & 0.94\textsuperscript{12} & 0.75\textsuperscript{13} & 0.83\textsuperscript{14} & -5.90 \\ \hline
				K & Kakangari & 1.02\textsuperscript{15} & 1.04\textsuperscript{15} & 0.98\textsuperscript{15} & 1.52  \\ \hline
				O & various & 0.93\textsuperscript{16} & 0.78\textsuperscript{16} & 0.88\textsuperscript{16} & -1.69  \\\hline
			\end{tabularx}
		\end{center}
	\end{table}
	
	\subsection{Numerical methods}
	
	In order to solve for the transport of solids and any possible chemical mixing in a time-dependent manner we must proceed numerically. Our numerical method integrates the evolution of \textbf{the gas} (\ref{1D gas evolution equation})  and \textbf{dust} (\ref{1D dust evolution equation}) \textbf{surface densities} explicitly on a staggered non-uniform mesh using finite volume operators. It is identical to that described in \citet{Owen2014} and has previously been used for astrophysical and cosmochemical applications by \citet{OwenArmitage2014,OwenJacquet2015}. The method is second-order-accurate in space and first-order-accurate in time. For the advection term in Equation~(\ref{1D dust evolution equation}) we use second-order reconstructions and Van-Leer limiters at cell boundaries. At the inner boundary we adopt zero torque boundary conditions for the gas and free-outflow boundary conditions for the solids. In the simulations with a steady disk profile we have constant mass-flux boundary conditions at the outer edge, in both the gas and the dust components. In the simulations with an evolving disk we apply at the outer edge a zero-torque boundary for the gas and free-outflow for the solids, but we note that in these simulations the outer boundary is chosen to be at a distance such that it does not affect the evolution. 
	
	Unless otherwise noted, the simulations were run with 800 grid elements spaced uniformly in $R^{1/8}$ between $R_\mathrm{in}=3\times10^{11}\,\mathrm{cm}$ and $R_\mathrm{out} = 3\times10^{17}\,\mathrm{cm}$.

	\section{Evolution of a single pair of chondrule and dust populations}
	\label{Dust top hat model}

	In this section, we investigate steady disk models with a fixed, initial pair of chondrule and dust populations, that is only consider a single CFE.

	\subsection{Complementary start}

	We first consider the case where chondrules and dust are cogenetic and originally in complementary proportions. We assume that the initial populations are within a ``top hat" of width $L$ centred at radius $R_\mathrm{centre}$. Within the limits $R=R_\mathrm{centre}-L/2$ and $R=R_\mathrm{centre}+L/2$, the chondrule and dust surface densities have
	$\Sigma_{\mathrm{ch}}=\Sigma_{\mathrm{d}}=0.01\cdot \Sigma_g$ (but note that the complementarity parameter $\zeta$ at any time is independent of the initial solid/gas ratios chosen). We evolve the populations over $2\, \mathrm{Myr}$ for values of $R_\mathrm{centre}$, $L/R_\mathrm{centre}$, and $\dot{M}$ ranging between $0.3$--$30 \, \mathrm{AU}$, $0.05$--$1.5$, and $10^{-10}$--$10^{-6}\,\mathrm{M}_{\odot} \mathrm{yr}^{-1}$, respectively. 
	
\begin{figure*} \includegraphics[width=\textwidth]{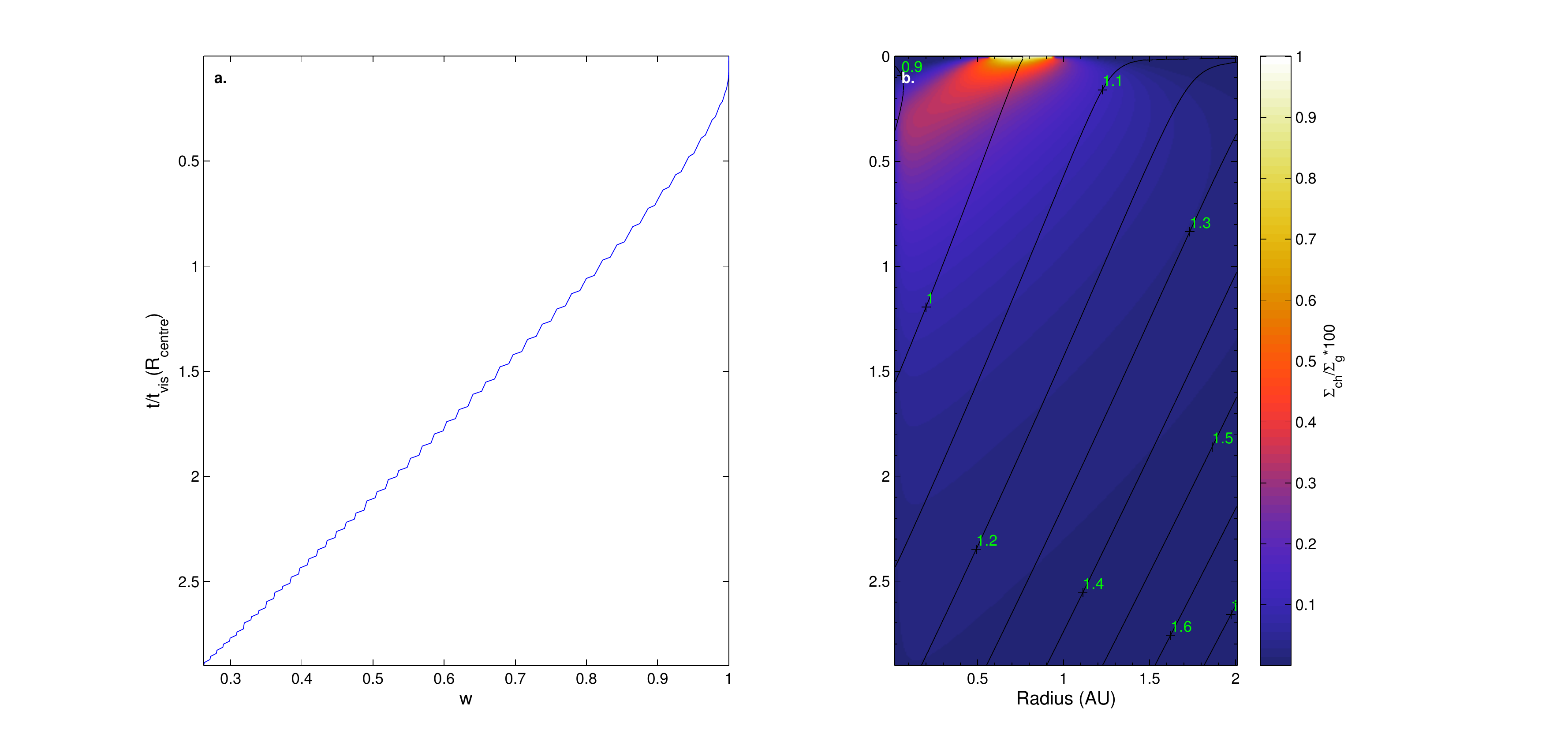}
		\caption{\textbf{a.} Non-dimensionalized time $t/t_{\mathrm{vis}}(R_\mathrm{centre})$ versus complementary fraction $w$, for $R_\mathrm{centre}=0.75\, \mathrm{AU}$, $L/R_\mathrm{centre}=0.5$, and $\dot{M}=1.36\times 10^{-8} \,\mathrm{M}_{\odot} \mathrm{yr}^{-1}$. The mass fraction of complementary chondrules is seen to decrease monotonically from 1, with $t_\mathrm{half} \approx 2 t_{\mathrm{vis}}$. \textbf{b.} Space-time plot of the chondrule population evolution for the same parameters. The plot is coloured according to chondrule mass fraction, overlain on top of which are contours of constant $\zeta$. Both plots have a common y-axis, with non-dimensionalized time increasing from top to bottom on the graphs.}
		\label{two-pop-sample} 
	\end{figure*}

	Over time, the chondrule population tends to drift inward and its width grows larger than $L$, exemplified by Figure \ref{two-pop-sample}b. Its inward drift being faster than that of the dust, decoupling between the two originally complementary populations progressively occurs. To make this quantitative, we calculated the fraction $w$ (the ``complementary fraction") of the chondrule mass within the simulation domain for which the complementarity parameter (here independent of any assumption on the composition of chondrules relative to matrix) is in the conventional complementary range (0.7--1.3; see Section \ref{section-complementarity}). By construction, the system is completely complementary ($w=1$) at time $t=0$, but $w$ monotonically decreases with time (Figure \ref{two-pop-sample}a). We call $t_{\rm half}$ the time at which $w=0.5$, which serves as a measure of the chondrule/dust decoherence timescale \citep{Jacquetetal2012S}.
	
\begin{figure}\includegraphics[width=\columnwidth]{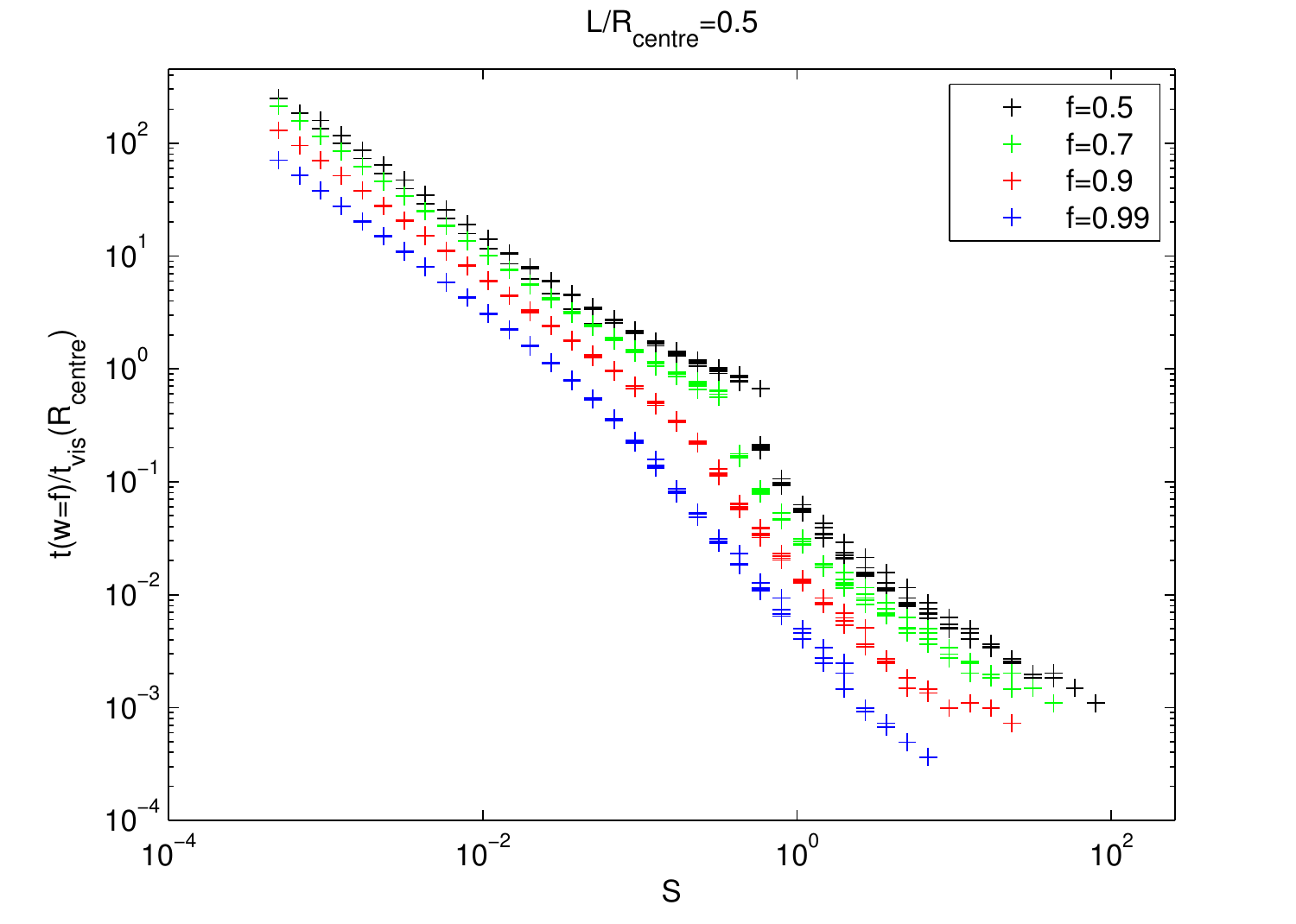}
		\caption{Log-log plot of $t(w=f)/t_{\mathrm{vis}}(R_{\rm centre})$ versus $S$ for varying $R_\mathrm{centre}$ and $\dot{M}$, evaluated at $R_\mathrm{centre}$, $\dot{M}$, and various values of $f$. \textbf{The value of} $w$ drops below larger values of $f$ in shorter amounts of time, but the slopes do not rely heavily on the chosen value of $f$ for any value of $S$.}
		\label{varying-w}
	\end{figure}

\begin{figure}\includegraphics[width=\columnwidth]{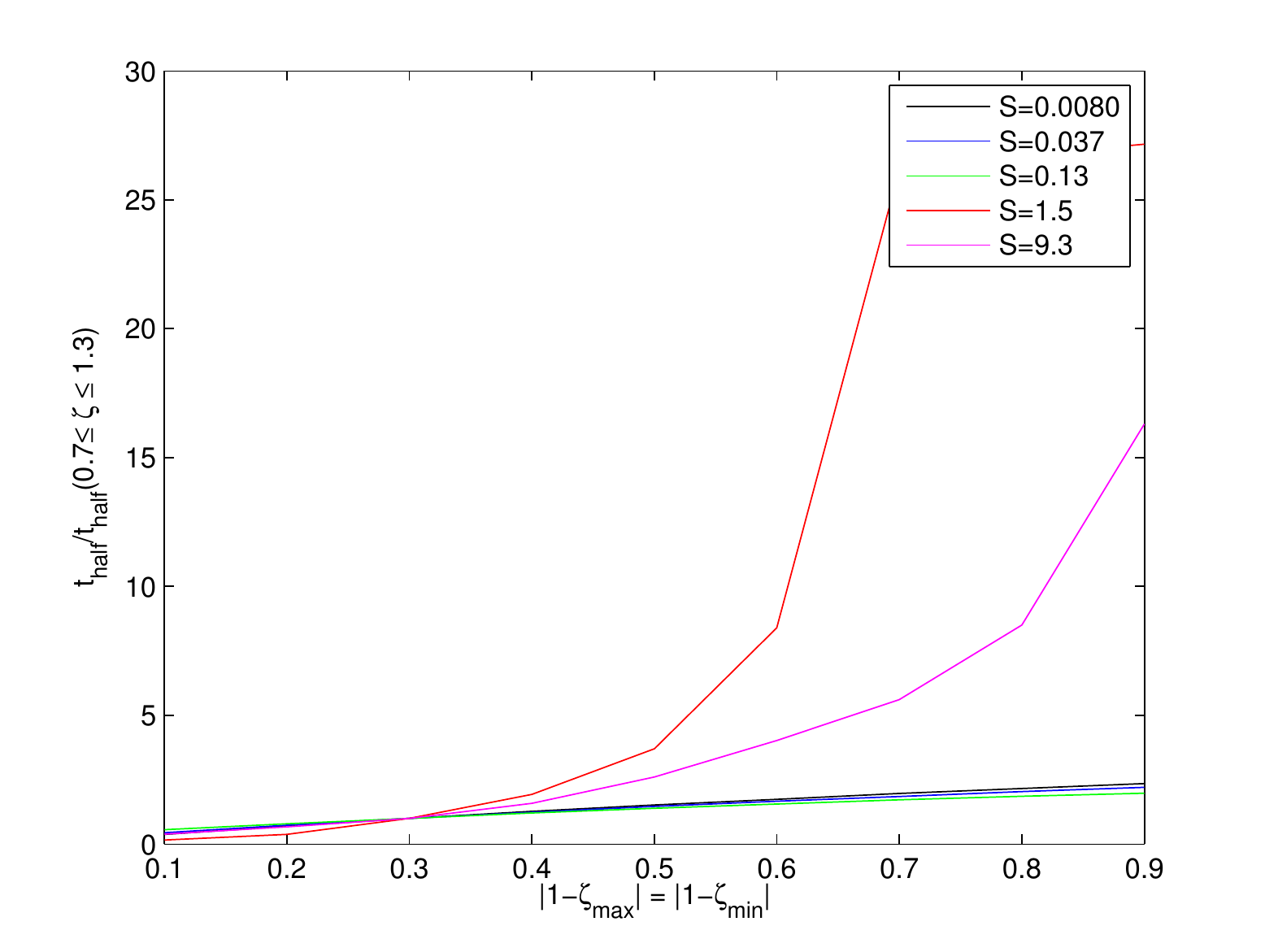}
		\caption{Plot of $t_\mathrm{half}$ for various definitions of complementarity, all with $L/R_\mathrm{centre}=0.1$. Values of $t_\mathrm{half}$ are normalized to the value corresponding to the standard complementary range ($0.7\leq\zeta\leq1.3$) in the paper.
		}
		\label{varying-zeta}
	\end{figure}

	In Figure \ref{varying-w}, we plot $t_{\rm half}/t_{\rm vis}(R_{\rm centre})$ versus $S(R_{\rm centre})$ (evaluated for chondrules). Under this normalization, the points for a given $L/R_{\rm centre}$ value collapse on a single line. This is due to the fact that the solid evolution equations (\ref{1D dust evolution equation}) can be non-dimensionalized in terms of only $R/R_{\rm centre}$, $t/t_{\mathrm{vis}}(R_{\rm centre})$, and $S(R_{\rm centre})$ for a steady disk. Figures \ref{varying-w} and \ref{varying-zeta} show how this behaviour is modified if the threshold complementary fraction or the range of $\zeta$ deemed complementary are changed, respectively.
	
	Several regimes can be distinguished depending on the value of $S$: 
	
	For very large $S$, specifically $S>R_{\rm centre}/L$, gas-solid drift is dominant and diffusion hardly affects the chondrule and dust profiles, which are essentially translated inward. $w=0.5$ then corresponds to the case where the chondrule population is shifted by $L/2$ relative to the dust population, which occurs for the time
	\begin{equation}
		\begin{aligned}
			t_\mathrm{half}=\frac{L/2}{\left| v_{\mathrm{drift}}\right|}&=\frac{L\rho}{2\tau}\left| \frac{ \partial P}{\partial R}\right|^{-1}\\
			&\mathbf{=\frac{L}{2}\frac{t_\mathrm{vis}(R_\mathrm{centre})}{S(R_\mathrm{centre})}\left|\frac{\partial \ln P}{\partial \ln R}\right|^{-1}}
		\end{aligned}
	\end{equation}
	(neglecting the drift of the dust grains relative to the gas), yielding (for a steady, constant $\alpha$ disk\textbf{; i.e., $\mathbf{P\propto R^{3-q/2}}$})
	\begin{equation}
	\frac{t_\mathrm{half}}{t_{\mathrm{vis}}(R_\mathrm{centre})}= \left(\frac{L}{R_\mathrm{centre}}\right)\left(6-q\right)^{-1}S\mathbf{(R_\mathrm{centre})}^{-1}.
	\label{t-half large S}
	\end{equation}
	
	For $1\leq S\leq R_\mathrm{centre}/L$, diffusional widening of the initial populations cannot be ignored, and in fact rapidly dominates over the initial width $L$, delaying the decoherence between chondrules and dust \citep{Jacquetetal2012S}. We show in appendix \ref{Jacquet regime} that 
	\begin{equation}
	t_\mathrm{half}=\frac{2D}{2.297^2}\left(\frac{\rho}{\tau\cdot \partial P/\partial R}\right)^2.
	\end{equation}
	This \textbf{leads to the relation}

	\begin{eqnarray}
	\frac{t_\mathrm{half}}{t_\mathrm{vis}(R_\mathrm {centre})}&=&\frac{2}{2.297^2}\frac{1}{Sc S\mathbf{(R_{\rm centre})}^2}\left( \frac{ \partial \ln P}{\partial \ln R}\right)^{-2}\nonumber\\
	&=&\frac{0.38 Sc^{-1}}{\left(\left(3-q/2\right) S\mathbf{(R_{\rm centre})}\right)^2}
	\label{t-half middle S}.
	\end{eqnarray}

	For $S\ll 1$, both chondrules and dust should follow the gas and be advected to the disk's centre by time $\sim t_{\mathrm{vis}}$. However, complementarity may linger for even longer timescales while part of the chondrules and dust are diffused outward (as $w$ is indeed defined in terms of the \textit{remaining} chondrule population). We may estimate that complementarity will be lost when diffusion will have reached the $S_R=S\cdot Sc=1$ line, which is the cutoff for outward diffusion for the chondrules \textbf{but not for the dust} \citep{Jacquetetal2012S}, that is on a timescale 
	\begin{equation}
	t_\mathrm{half} \sim t_{\mathrm{vis}}(S_R=1)=t_{\mathrm{vis}}(R_\mathrm{centre})\left(\frac{R(S_R=1)}{R_\mathrm{centre}}\right)^{q+1/2}
	\end{equation}
	and so
	\begin{eqnarray}
	\frac{t_\mathrm{half}}{t_{\mathrm{vis}}(R_\mathrm{centre})}&\sim &\left(S_R(R_{\rm centre})^{-1/(3/2-q)}\right)^{q+1/2}\nonumber\\
	&=&\left(S(R_{\rm centre})\cdot Sc\right)^{1-2/(3/2-q)}
	\label{t-half small S}.
	\end{eqnarray}
	For our $q$ value, this corresponds to a $S^{-1}$ dependence as observed. Additional runs with alternative $q$ values change the slope of the log-log plot in Figure \ref{varying-w} in accordance with this prediction. Clearly, for $S\ll 1$, complementarity can be maintained for timescales as large as the viscous timescale, as advocated by \citet{Jacquetetal2012S}.

\begin{figure*} \includegraphics[width=0.85\textwidth]{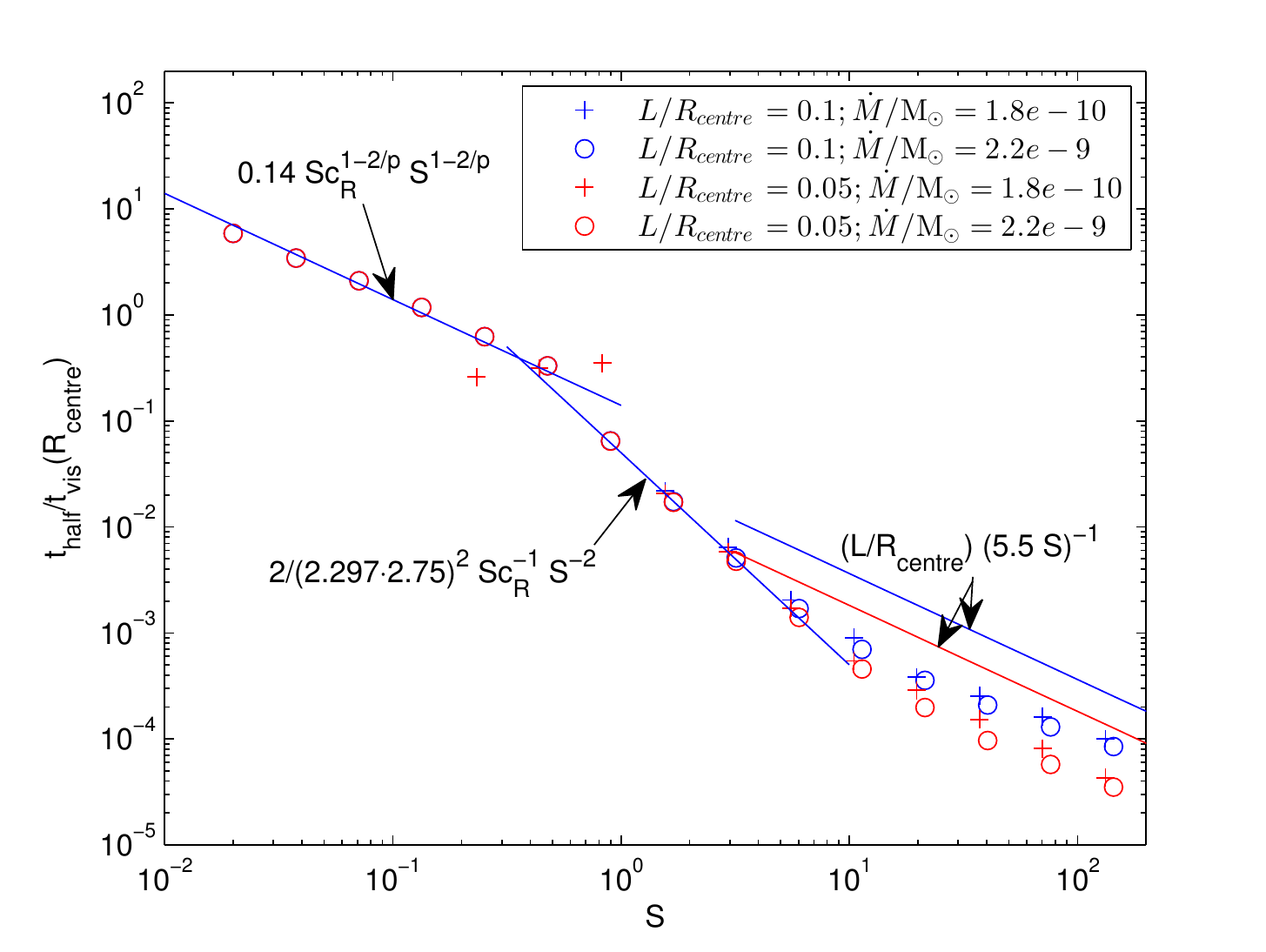}
		\caption{Log-log plot of $t_\mathrm{half}/t_{\mathrm{vis}}$ versus $S$ for varying $R_\mathrm{centre}$, $\dot{M}$, and $L/R_\mathrm{centre}$. Both are evaluated at $R_\mathrm{centre}$ and $\dot{M}$, and overplotted \textbf{in solid blue and red lines are} the analytical predictions. Red and blue points correspond to $L/R_\mathrm{centre}=0.05$ and $L/R_\mathrm{centre}=0.1$, respectively; plus signs and circles correspond to $\dot{M}=1.85\times10^{-10}\,\mathrm{M}_\odot \mathrm{yr}^{-1}$ and $\dot{M}=2.15\times10^{-9}\,\mathrm{M}_\odot \mathrm{yr}^{-1}$, respectively. The blue and red points appear as one red point for $S\leq 1$, as they completely overlap.
		}
		\label{thalf-high-res}
	\end{figure*}

	We have re-run the trials with only two $\dot{M}$ values, a wider range of $R_\mathrm{centre}$ values, and 16 times the spatial resolution, to validate convergence to the analytical results (Figure \ref{thalf-high-res}). The $-1$ slope from (\ref{t-half small S}) and exact solution 
	taking $q=0.5$ for (\ref{t-half middle S}) robustly match the data, but the predicted behaviour (\ref{t-half large S}) at large $S$ is off by a factor of about $2/3$. The latter may be attributed to rapid diffusion at the sharp top hat edges with which the populations are initialised, as it improved slowly with increased spatial resolution. Setting $Sc=10^8$ to nullify the effects of diffusion for a few trials with $S\gg 1$ resulted in agreement with the analytical predictions to within $2\%$. 
	
	\subsection{``Two-component'' start}
	
\begin{figure*}\includegraphics[width=\textwidth, height=10cm, keepaspectratio]{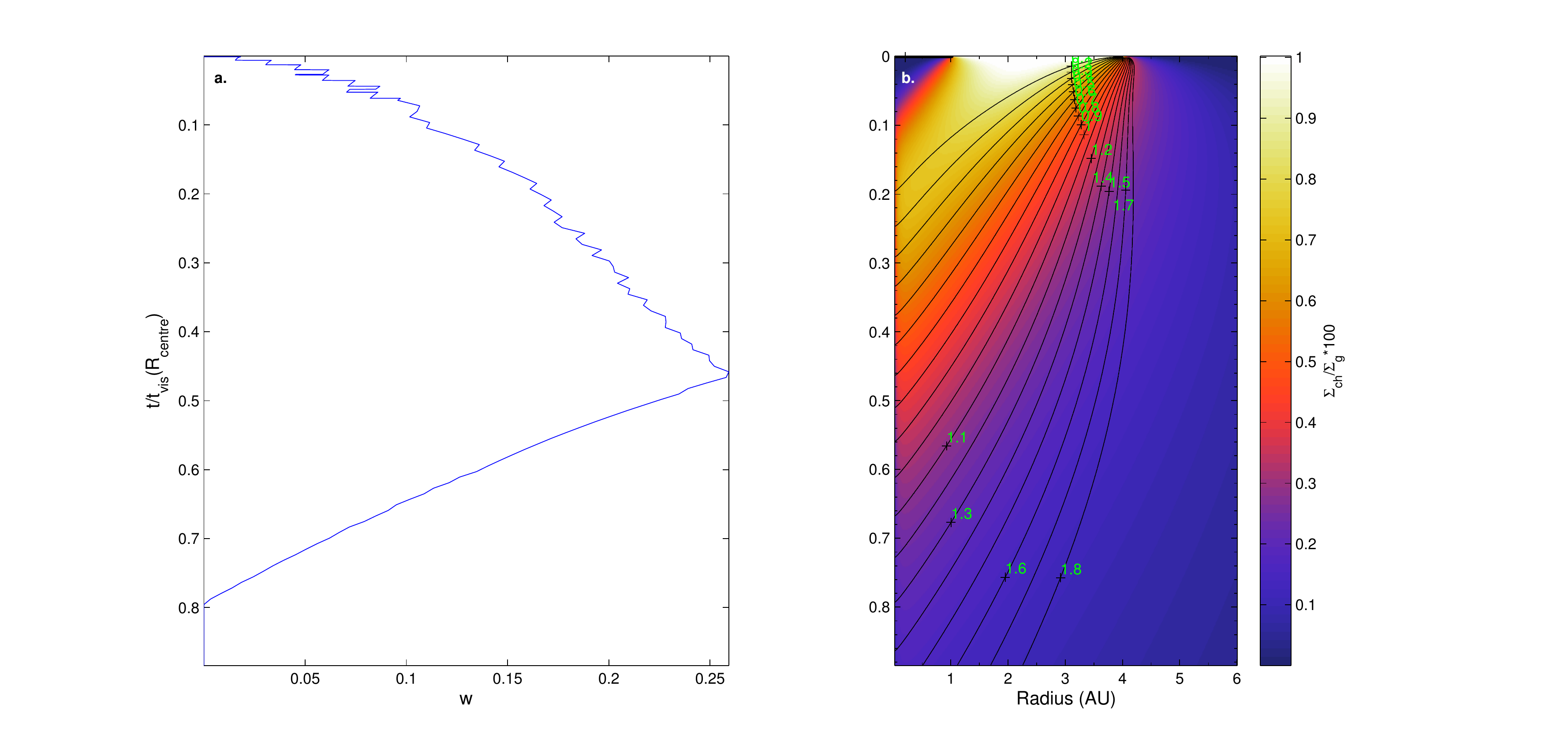}
		\caption{\textbf{a.} Non-dimensionalized time $t/t_{\mathrm{vis}}(R_\mathrm{centre})$ versus complementary fraction $w$, for $R_\mathrm{centre}=4.1 \,\mathrm{AU}$, $L/R_\mathrm{centre}=1.5$, and $\dot{M}=1\times 10^{-7} \,\mathrm{M}_{\odot} \mathrm{yr}^{-1}$. The mass fraction of complementary chondrules is seen to steadily increase from $w=0$ to $w\approx 0.25$ before decreasing back to $0$, with no $t(w\geq 0.5)$. \textbf{b.} Space-time plot of the chondrule population evolution for the same parameters. This plot is coloured according to chondrule mass fraction, overlain on top of which are contours of constant $\zeta$. Both plots have a common y-axis, with non-dimensionalized time increasing from top to bottom of the graphs.}
		\label{sample-two-component}
	\end{figure*}

	As a comparison, we have run simulations with spatially separated chondrules and dust populations, with the former and the latter initialized within the inner and outer halves of the top hat, respectively, with $L/R_\mathrm{centre}=1$ and $1.5$. This mimics a ``two-component'' picture where chondrules are derived from the inner disk, whereas the matrix represents pristine outer disk matter. Here, chondrules and dust are assumed to initially be in ``global'' complementary proportions\footnote{It should be noted that this may not exactly reflect the two-component model discussed in cosmochemistry, since the latter often assumes the matrix to be solar \citep[e.g.][]{Zandaetal2012}, although \citet{Gonzalez2014} suggests CI chondrites (which are used to represent solar abundances) are systematically volatile-enriched relative to the Sun. Here, our point is merely to evaluate how difficult complementarity is to obtain from non-cogenetic matrix and chondrules.}. Since they are spatially distinct, the initial complementary fraction $w$ as defined above is zero. With the passage of time, chondrule particles diffuse outward and some dust particles drift inward, thus $w$ increases and reaches a maximum of $w=0.5$-$0.6$ at time $t\leq t_\mathrm{half}$, then decreases back to $0$ as chondrules drift past the inner boundary (e.g. Figure \ref{sample-two-component}). Regardless of the values of $\dot{M}$ and $R_\mathrm{centre}$, the total amount of time with $w \geq 0.5$ never exceeds $2.3\times 10^{-5} t_{\mathrm{vis}}(R_{\rm centre})$. Complementarity, provided it is indeed a pristine feature of carbonaceous chondrites, is therefore much more difficult to obtain in a two-component picture vis-\`{a}-vis a single reservoir model. 
	
	\section{Continuous chondrule formation}
	\label{section-continuous-CFE}
	
	While the previous section considered a single initial CFE and the evolution of its products, we now investigate models with multiple epochs of chondrule formation. Following the results of the previous section, we shall consider that chondrule and matrix grain formation were not spatially distinguished, save for whichever fraction of an initial CI chondritic matrix component escaped CFEs until chondrite agglomeration. We first present our prescriptions for chondrule formation, then simulations for steady disks, and finally results for an evolving disk scenario.
	
	\subsection{Initial conditions and chondrule formation model}
	
	We consider initially two populations of CI chondritic composition (in terms of a fiducial element $X$) particles, one micron-sized (the dust) and another millimeter-sized (``aggregates"). Neither growth nor fragmentation are modelled. The solids' initial surface density profiles follow the static solution \textbf{Equation A.6} (for a static gas disk) derived by \citet{JacquetRobert2013}:
	\begin{equation}
	\frac{\Sigma_{\mathrm{d,agg}}}{\Sigma_\mathrm{g}}(R)=\epsilon\frac{3 Sc_R}{2}\exp(I(R)) \int_{R_{*}}^R\frac{\exp(-I(R^\prime))}{R^\prime}\,\mathop{\mathrm{d}R^\prime}
	\end{equation}
	with $\epsilon=0.01$ and
	\begin{equation}
	I(R)=\int_{R_{*}}^R\frac{v_R}{D_R}(R')\,\mathop{\mathrm{d}R'},
	\end{equation}
	\textbf{with the outer boundary conditions for both populations set to maintain the accretion rate (neglecting the diffusive flux)}
	\begin{equation}
	\dot{M}_{\mathrm{d,agg}}\equiv-2\pi R_\mathrm{out}\Sigma_{\mathrm{d,agg}}(R_\mathrm{out})v_{R;d,agg}(R_\mathrm{out})=\epsilon \dot{M}(R_{\rm out}).
	\end{equation}
	
	Chondrule formation is modelled with a chondrule production function ($g(R)$) defined such that, for any time interval $\mathop{\mathrm{d}t}$, a fraction $g \mathop{\mathrm{d}t}$ of the millimetre-sized bodies (originally only aggregates, but later also including previously formed chondrules) at any given location is converted into new chondrules\footnote{Other prescriptions were tried where part of the dust is also converted into chondrules but were not found to significantly differ from our standard ``conservative" prescription. For the sake of simplicity, given the arbitrariness of the chondrule recipe anyway, we shall solely focus on the latter.}. The same fraction of the existing dust becomes a new dust population. As a proxy for chemical exchange between dust and chondrules, the concentration of $X$ in the new chondrules and dust is fixed such that the local chondrule/dust ``partition coefficient"
	\begin{equation}
	\frac{[X]_\mathrm{ch}^*}{[X]_\mathrm{d}^*}=\beta
	\label{beta}
	\end {equation}	
	is a constant taken to be 1.37, an average calculated for CC data, with the additional constraint of conservation of bulk chemical abundance during the CFE. When assessing potential chondrite composition, the remaining primordial millimeter-sized aggregates at the location and time of consideration were counted among the matrix component.
	
	We chose $g$ to be zero outside of a prescribed chondrule forming region (CFR) between 0.5 AU and $R_{\rm CFE}$ (which was varied from 3 to 25 AU) and to be of the form
	\begin{equation}
	g(R)=\frac{A}{t_{\mathrm{vis}}(1\,\rm AU)} \times \left ( \frac{R}{1\,\mathrm{AU}}\right)^{-\delta}
	\end{equation}
	in the CFR. $A$ is a constant, corresponding to the fraction converted at 1 AU within one viscous timescale, and $\delta$ indicates the radial dependence of chondrule formation, ranging from $0$ (no dependence) to $2$.

	Numerically, the CFR was divided into several radial bins, and the CFEs were discretized to take place with a finite period $\mathrm{d}t$. For each CFE time and each radial bin, a (Eulerian) chondrule and a dust population (originally confined to the radial bin in question) are followed, that is are first created at the CFE time in question (with subtraction of the corresponding precursors from the previous populations), and then have their radial distributions updated for the remainder of the simulation following Equation (\ref{1D dust evolution equation}). For simplicity, since information from the exact provenance inside a radial bin is lost, a single (average) composition was assigned to each chondrule and dust population, corresponding to the average one resulting from Equation (\ref{beta}) and mass conservation.

	\subsection{Steady disk simulations}
	\label{Steady disk simulations}
	
	\begin{figure*}\includegraphics[width=\textwidth]{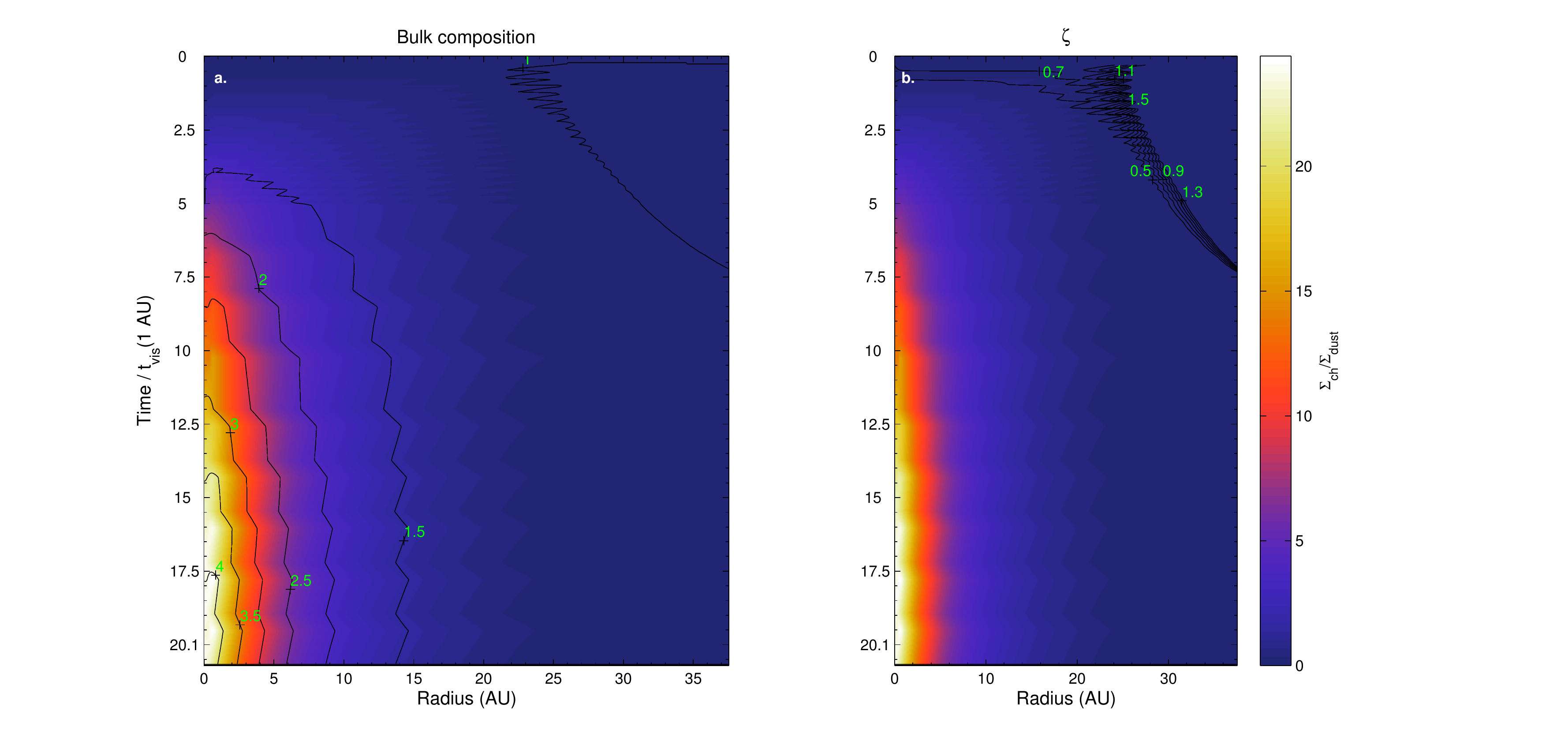}
		\caption{Space-time plot of the chondrule-to-dust surface density ratio, overplotted with contours of constant \textbf{a.} CI-normalized bulk composition and \textbf{b.} complementarity parameter $\zeta$. This simulation has $\dot{M}=10^{-8}\,\mathrm{M}_\odot\mathrm{yr}^{-1}$, $\delta=0$, $R_\mathrm{CFE}=25\,\mathrm{AU}$, and $A=0.75$.}
		\label{small-delta ex2}
	\end{figure*}
	
	\begin{figure*}\includegraphics[width=\textwidth]{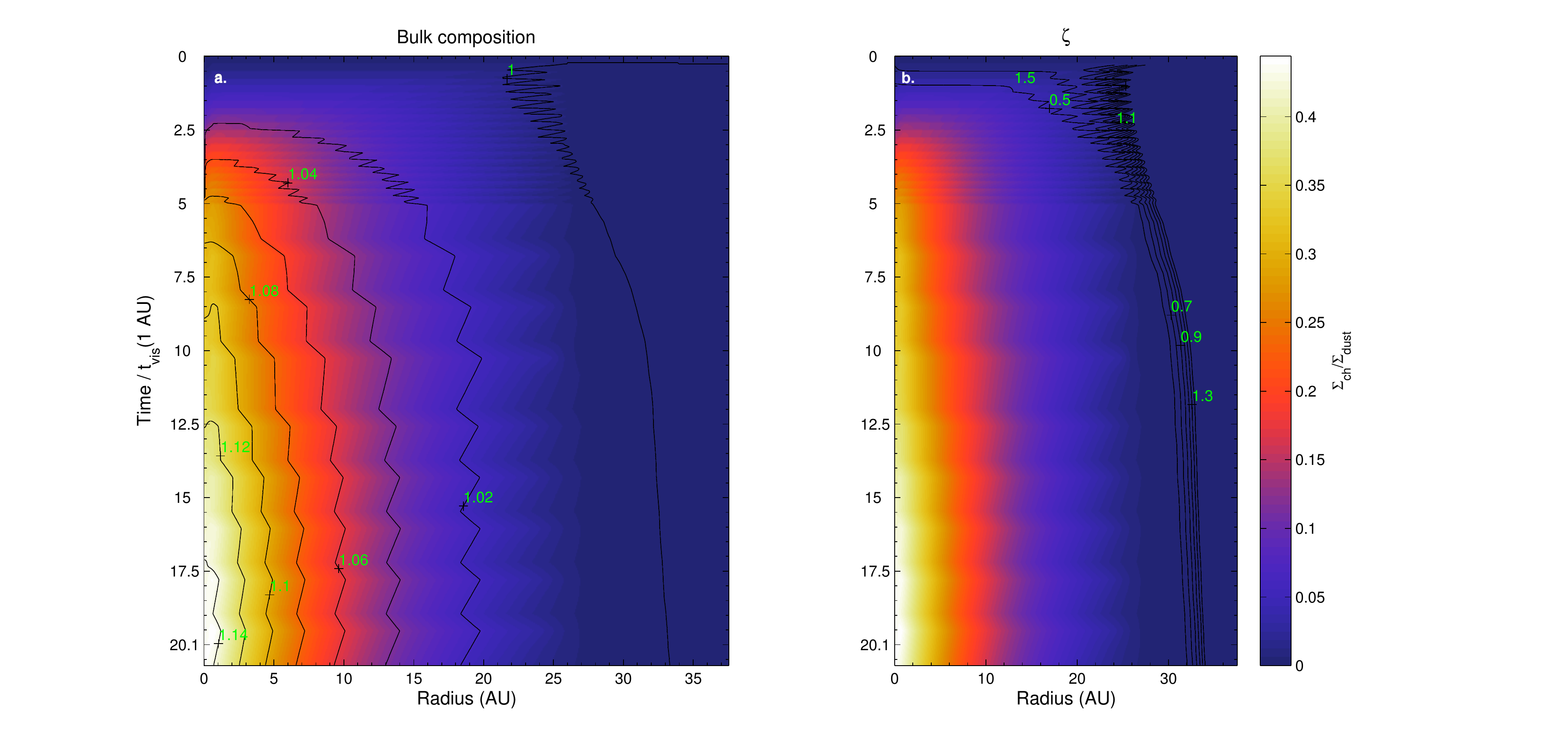}
		\caption{Same as Figure \ref{small-delta ex2}, but with a lower $A=0.075$. $\Sigma_{\mathrm{ch}}/\Sigma_{\mathrm{d}}$ decreases accordingly by one order of magnitude, CI-normalized bulk composition remains within $10\%$ of $1$ at all times, and the complementary range of $\zeta$ moves to smaller radii (and remains constant).}
		\label{small-delta ex3}
	\end{figure*}
	
	\begin{figure*}\includegraphics[width=\textwidth]{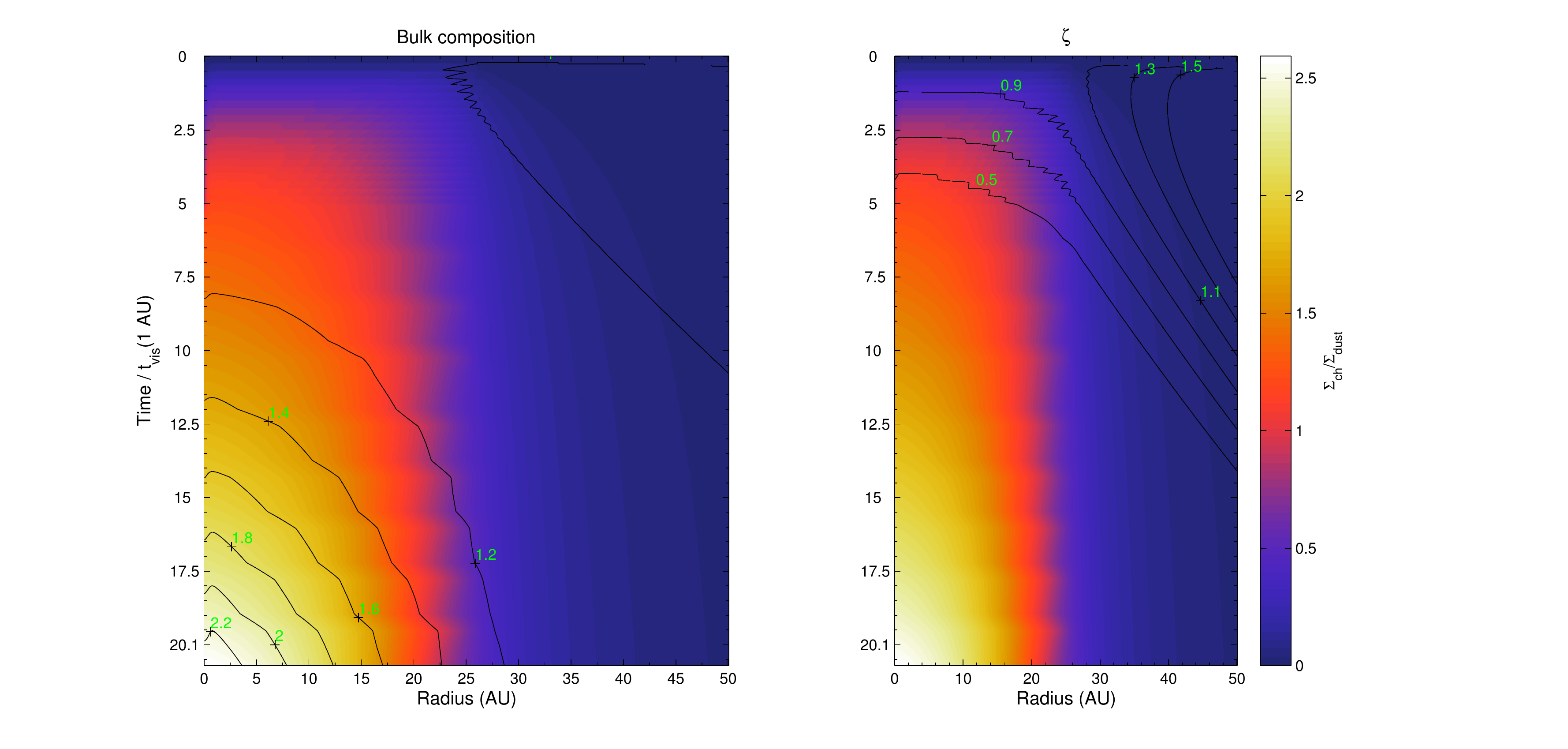}
		\caption{Same as Figure \ref{small-delta ex2}, but with a higher $\dot{M}=10^{-7}\,\mathrm{M}_\odot\mathrm{yr}^{-1}$. In contrast to Figures \ref{small-delta ex2} and \ref{small-delta ex3}, chondrules occupy a portion of the disk out to $R\approx20\,\mathrm{AU}$.}
		\label{small-delta ex4}
	\end{figure*}
	
	We have run simulations using steady gas disks with $\dot{M}$ from $10^{-9}\,\mathrm{M}_{\odot} \mathrm{yr}^{-1}$ to $10^{-7}\,\mathrm{M}_{\odot} \mathrm{yr}^{-1}$ for $0.84\,\mathrm{Myr}\approx21 t_{\mathrm{vis}}(1\,\mathrm{AU})$. We varied $\delta$ (0 and 2), $R_\mathrm{CFE}$ (3, 10, and 25 AU), and $A$ (0.075 and 0.75). 12 radial bins were used with the CFE period set at $\mathop{\mathrm{d}t}=0.01\mathrm{Myr}$.
	
	\begin{figure*}\includegraphics[width=\textwidth]{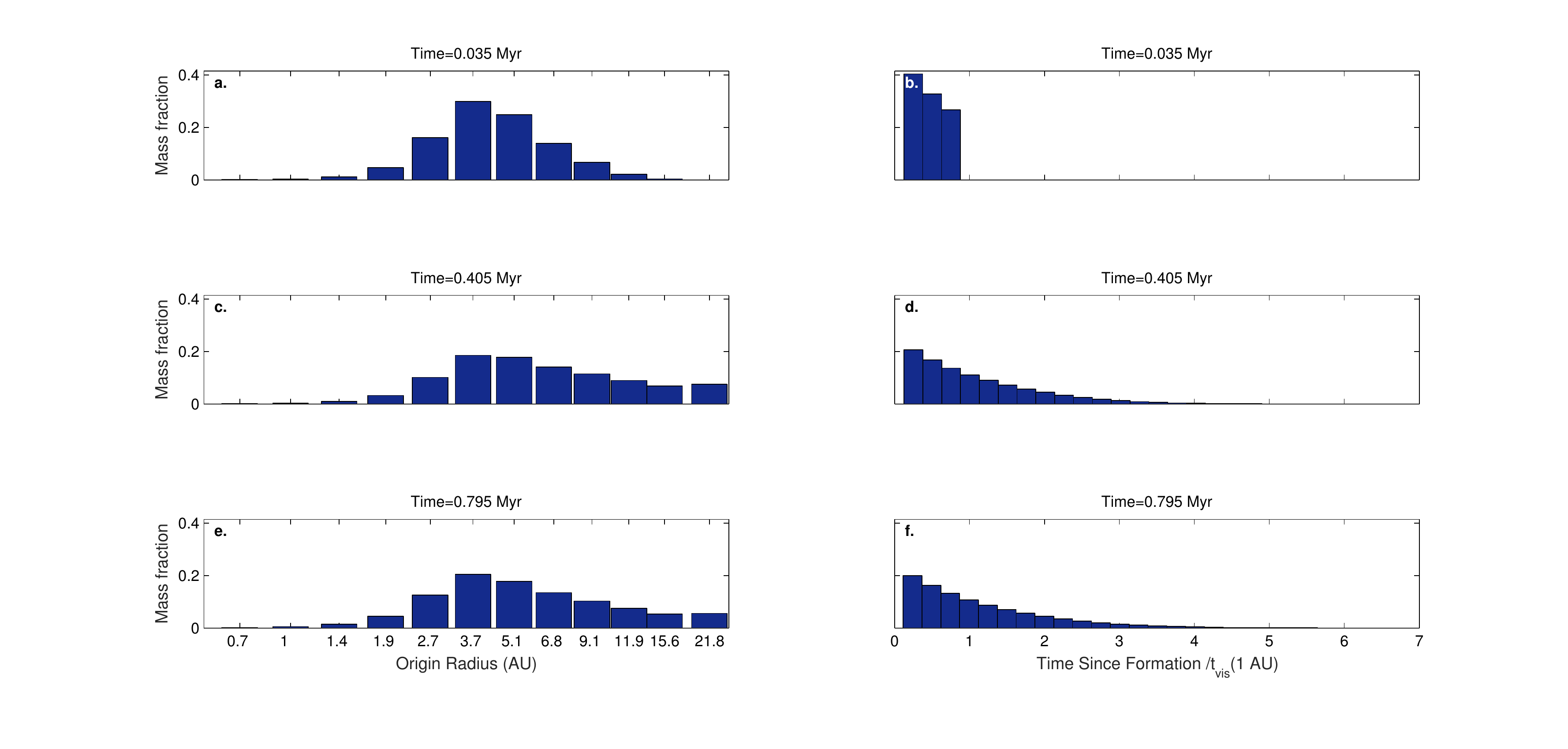}
		\caption{\textbf{Origin distributions of chondrules found at} $R=3\,\mathrm{AU}$, in terms of mass fraction, after \textbf{a.} \& \textbf{b.} $0.035\,\mathrm{Myr}$ ($0.9t_\mathrm{vis}(1\,\mathrm{AU})$), \textbf{c.} \& \textbf{d.} $0.405\,\mathrm{Myr}$ ($10.2 t_\mathrm{vis}(1\,\mathrm{AU})$), and \textbf{e.} \& \textbf{f.} $0.795\,\mathrm{Myr}$ ($20.0 t_\mathrm{vis}(1\,\mathrm{AU})$), each of which are halfway between CFEs. Here, $\dot{M}=10^{-8}\,\mathrm{M}_\odot\mathrm{yr}^{-1}$, $\delta=0$, $R_\mathrm{CFE}=25\,\mathrm{AU}$, and $A=0.75$ are used. Note that the age is defined relative to accretion time, not to the start of the simulation. Origin radius is taken as the middle grid cell of each chondrule formation bin. Chondrules are predominantly seen to have formed at radii close to their current radius, with an emphasis towards larger radii, and their age distribution falls off exponentially with time since formation. These distributions achieve a steady state by $t\approx10t_\mathrm{vis}(1\,\mathrm{AU})$.}
		\label{steady-histograms}
	\end{figure*}

	Figures \ref{small-delta ex2}, \ref{small-delta ex3}, and \ref{small-delta ex4} show space-time diagrams of the chondrule/dust ratio, with contours of the bulk composition as well as the complementarity parameter, for the case $A=0.75$ (0.075 in Figure \ref{small-delta ex3}), $\delta=0$, $R_{\rm CFE}=25\:\rm AU$, and $\dot{M}=10^{-8}\,\mathrm{M}_\odot\mathrm{yr}^{-1}$ ($10^{-7}\,\mathrm{M}_\odot\mathrm{yr}^{-1}$ in Figure \ref{small-delta ex4}). We witness the initial increase of the chondrule abundance (initially zero) until some steady-state is reached with their loss by advection and drift into the Sun. 
	A radial gradient of the chondrule/dust ratio sets in, in part due to the limited extent of the CFR, but here mainly controlled by the differential drift between chondrules and dust (more or less coupled with the gas). As we mentioned previously, this starts to have an effect when $S\sim 1$ for the chondrules, corresponding to (using Equation (\ref{S steady})):
	\begin{equation}
	R_{\rm S=1}= 5\,\mathrm{AU}\left(\frac{\dot{M}}{10^{-8}\:\rm M_\odot/yr}\frac{0.04}{H/R(1\:\rm AU)}\frac{0.1\:\rm g/cm^2}{\rho_sa}\right)^{1/(3/2-q)} .
	\end{equation}
	Indeed, as $\dot{M}$ decreases (compare e.g. Figures \ref{small-delta ex4} to \ref{small-delta ex3}), the chondrules become more and more concentrated in the inner disk. Compositionally, this results in a more chondrule-enriched innermost region. Beyond this in the CFR is a chondrule-depleted region, which shows a \textit{slightly} subsolar composition in term of our fiducial ``chondrule-loving" element $X$ for the lowest $\dot{M}=10^{-9}$~M$_\odot$/yr, with yet further out a chondritic composition dominated by the unprocessed dust/aggregates. Only for the highest $\dot{M}=10^{-7}$~M$_\odot$/yr (Figure \ref{small-delta ex4}) do we see complementarity linger for a few $t_{\rm vis}$, in accordance with the previous section.
	
	For smaller $A$ values, the steady-state chondrule/dust ratio is lower and the departures from solar compositions are thus smaller. A very rough quantification of the production/loss balance is to consider that the net production of chondrules $g\Sigma_{\rm dust+agg}$ in the inner disk is balanced by the loss $-\Sigma_{\rm ch}(1+S)/t_{\rm vis}$ (advection+drift) in steady state, hence $\Sigma_{\rm ch}/\Sigma_{\rm dust+agg}\sim gt_{\rm vis}/(1+S)= AR_{\rm AU}^{3/2-q-\delta}/(1+S)$ (compare Figures \ref{small-delta ex2} and \ref{small-delta ex3}).

	Figure \ref{steady-histograms} shows the age and location of origin of chondrules found at $3\,\mathrm{AU}$ after $0.04$, $0.4$, and $0.8\,\mathrm{Myr}$ (midway between the two closest CFEs). From these histograms, we see the intuitive result that most of the chondrules at a given radius were formed in spatial bins near that radius, with a greater emphasis on the bins beyond it due to inward advection. The ages of the chondrules at this radius are mostly within $t_{\mathrm{vis}}(3\,\mathrm{AU})\approx0.12\,\mathrm{Myr}$, with a trend toward recent formation. The distributions are nearly identical at the later two times. This implies that the continuous CFEs lead to a constant distribution of spatiotemporal origin of chondrules at a given radius within $0.4\,\mathrm{Myr}\approx 3t_{\mathrm{vis}}(3\,\mathrm{AU})$, due to the continuous replenishment of chondrule material. Let us now study how the spatial and temporal sources of these depend on model parameters.
	
\begin{figure*}\includegraphics[width=\textwidth]{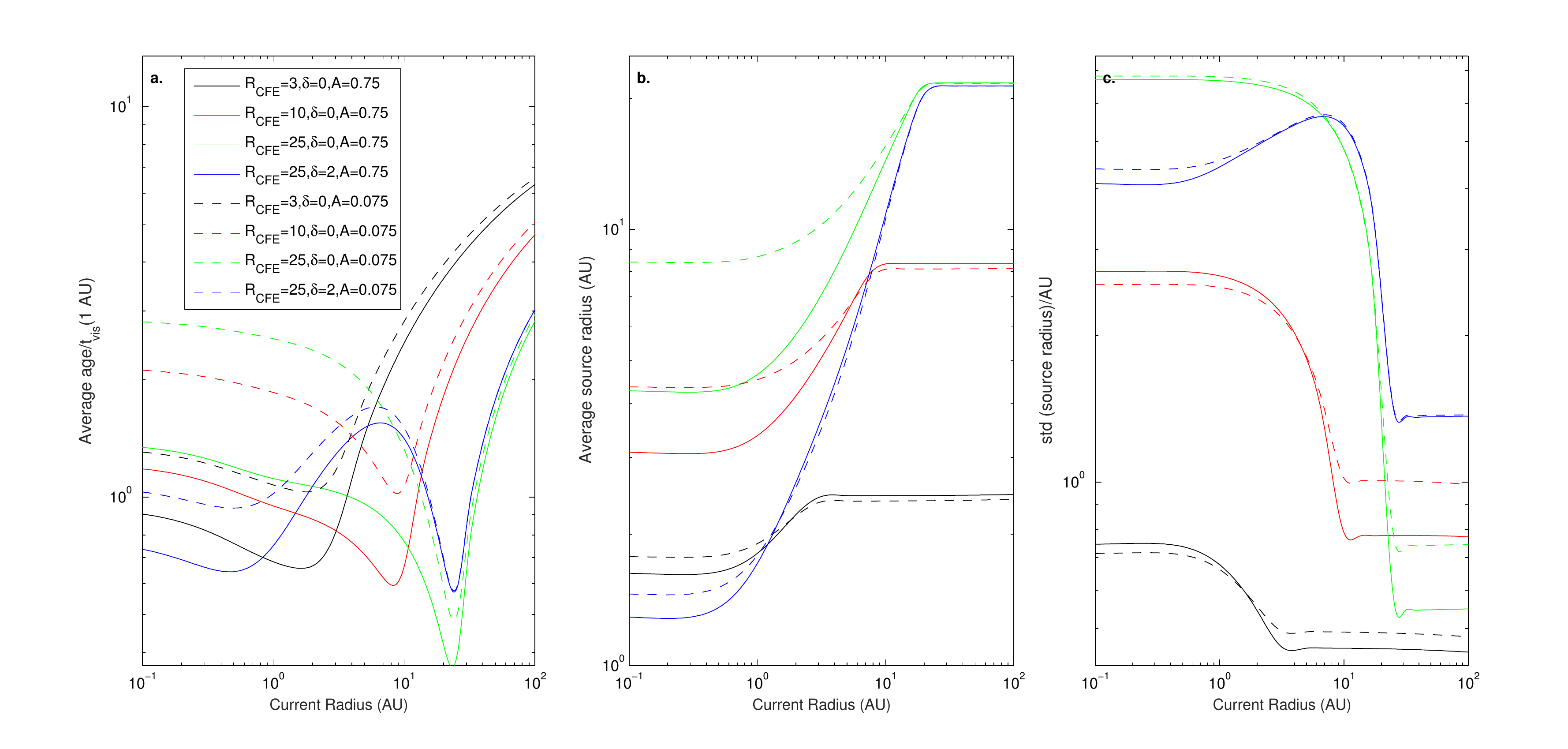}
		\caption{Chondrule origins after $0.835\,\mathrm{Myr}$, plotted versus $R$, with $\dot{M}$ restricted to $10^{-8}\,\mathrm{M}_\odot\mathrm{yr}^{-1}$. \textbf{a.} Average age normalized to $t_{\mathrm{vis}}$ is plotted. Age is seen to decrease until the end of the CFR, then rapidly increase until the limit of outward diffusion. For $\delta=2$, chondrule age grows through the middle and end of the CFR. Increasing $A$ simply reduces average chondrule age. \textbf{b.} Average source radius (AU) is plotted. This is constant for small $R$, increases with $R$ across the CFR, then plateaus. Increasing $R_\mathrm{CFE}$ increases average source radius, and decreasing $A$ increases source radius by a constant factor for radii smaller than the CFR. \textbf{c.} Standard deviation of source radius (AU) is plotted. The width of the radial distribution is seen to be constant for small $R$, decrease across the CFR, and then plateau, for $\delta=0$. For $\delta=2$, the distribution of source radii is more narrow at small $R$.}
		\label{comparing-A-combined}
	\end{figure*}

	Figure \ref{comparing-A-combined}a shows a plot of the average chondrule age as a function of $R$ (i.e., computed at each radial point) midway between the two final CFEs, for $\dot{M}=10^{-8}\,\mathrm{M}_{\odot} \mathrm{yr}^{-1}$. The average age is seen to be minimum within the CFR, which reflects the dominance of locally, and generally recently produced, chondrules. It steadily increases beyond $R_{\rm CFE}$, owing to the increasing transport time needed, until the limit of outward diffusion, where the minute amount of particles are homogenized in terms of spatiotemporal origin due to numerical effects. The $\delta=2$ case shows a more complicated behaviour, with two local minima, presumably because chondrules are essentially produced at small radii, amounting to a small effective outer boundary of the CFE. It shows an anticipated increase in age relative to the $\delta=0$ case, yet the few chondrules produced at large distances dominate further out and cause another ``local production" minimum.
	
	Figure \ref{comparing-A-combined}b shows the average source radius of chondrules as a function of radial location. Inside the inner boundary of the CFR, it shows a plateau value corresponding to chondrules advected dominantly from the inner regions of the CFR, before rising in the CFR, being comparable to, but larger than, the local radius, and then plateaus again outside of the CFR at $R\approx 0.85 R_\mathrm{CFE}$. For $R_{\rm CFE}=25\:\rm AU$, $\delta=2$ shows smaller average source radii than $\delta=0$, as expected from the greater concentration of chondrule production at shorter heliocentric distances. The plateau outside the CFR may be understood as follows: in steady state, the distribution of the integrated population of chondrules produced at a given heliocentric distance $R_{\rm source}$, which has no source (or sink) term there and zero value at infinity, is given by $\Sigma_i/\Sigma_g (R)\propto \left(\mathrm{exp}(-11S/4)/R^{3/2}\right)^{\rm Sc}$ \citep{Jacquetetal2012S}. Since the right-hand side is a fixed function of $R$, independent of $R_{\rm source}$, the \textit{relative} proportions of two integrated chondrule populations do not depend on $R$ and merely reflect the relative production rates as well as the relative ease of outward transport at the sources. The standard deviation of the source radius (Figure \ref{comparing-A-combined}c) decreases from about $R_{\rm CFE}/4$ inwards of the CFR to $\lesssim 1$ AU outwards, reflecting that chondrules produced close to the Sun are not transported efficiently outside the CFR (e.g., because the CFR ends beyond the $S=1$ line).

	Decreasing $A$ from 0.75 to 0.075 increases somewhat the average ages and radii, as expected from lower degrees of recycling (allowing longer average travels), but the effects are rather limited (order unity or less) especially at large heliocentric distances. This is because with such $A$ values $\leq 1$ we are mostly in the $g\ll t_{\rm vis}^{-1}$ regime where the behaviour of chondrules is dominated by transport, so that the age is rather limited by the viscous timescale of accretion to the Sun, and variations in $A$ do not affect the relative production efficiencies of the different radial locations. Of course, it may be noted, in the opposite regime of efficient recycling $g\gg t_{\rm vis}^{-1}$, the chondrules, which would be dominantly local, would have an average age asymptoting to $g^{-1}\propto A^{-1}$, with an exponential distribution.

\begin{figure*}\includegraphics[width=\textwidth]{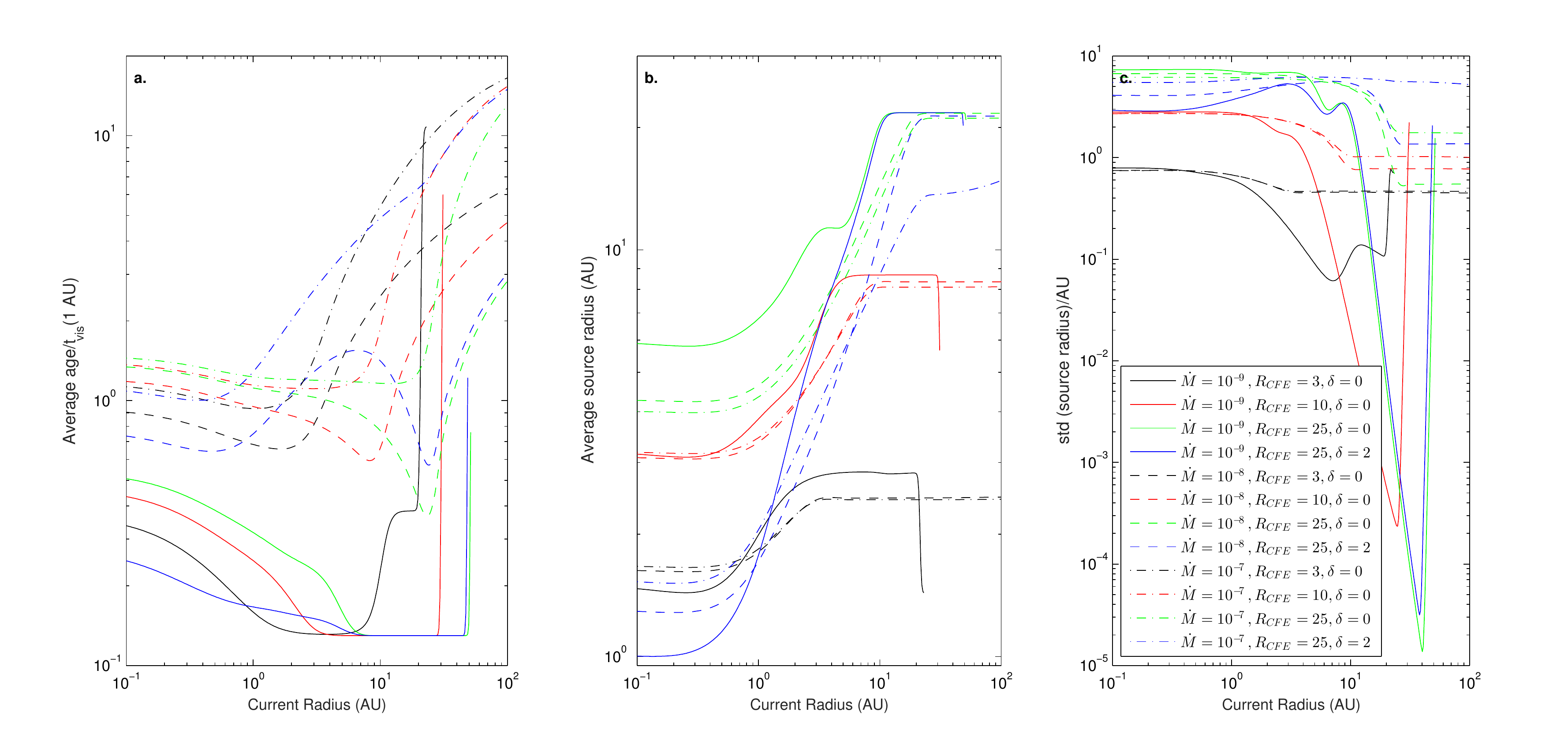}
		\caption{Chondrule origins after $0.835\,\mathrm{Myr}$, plotted versus $R$, with $A$ restricted to $0.75$. \textbf{a.} Average age normalized to $t_{\mathrm{vis}}$ is plotted. Increasing $\dot{M}$ shifts the behaviour of average age versus $R$ to higher average age. For low $\dot{M}$, average age achieves the minimum possible value and plateaus for sufficiently large $R$. \textbf{b.} Average source radius (AU) is plotted. Decreasing $\dot{M}$ similarly shifts the behaviour of average source radius versus $R$ to shorter values. \textbf{c.} Standard deviation of source radius (AU) is plotted. $\dot{M}$ does not seem to greatly affect the width of the source radius distributions at small $R$, and significantly decreases it as $R$ approaches $\sim30\,\mathrm{AU}$.}
		\label{combined-vs-S-bigger}
	\end{figure*}

	Decreasing $\dot{M}$ (Figure \ref{combined-vs-S-bigger}) tends to reduce the ages (down to half the numerical period for $\dot{M}=10^{-9}\:\rm M_\odot/yr$). This reflects weaker coupling of the chondrules with the gas ($S\propto\dot{M}^{-1}$), incurring more rapid radial drift to the Sun. The radial standard deviations decrease somewhat, as predicted by \citet{Jacquetetal2012S}, though this is a limited effect.

	\subsection{Evolving disk models}
	
	In our final set of calculations, we consider runs where the gas disk evolves. The surface density profile was initialized so as to match the zero time \citet{LyndenBellPringle1974} similarity solution 
	\begin{equation}
	\Sigma\left(R,\right)=\frac{M_\mathrm{disk}\left(0\right)}{2\pi R R_1}\exp\left(-\frac{R}{R_1}\right).
	\end{equation}
	$M_\mathrm{disk}\left(0\right)=0.07 \,\mathrm{M}_{\odot}$ and $R_1=18\,\mathrm{AU}$, corresponding to the disk's initial mass and scale radius respectively, were chosen so as to best fit observed accretion rates and disk fractions to the Lynden-Bell \& Pringle solution \citep{Owenetal2011}. These trials were run for $3.5$ Myr \citep[that is, the time at which the disk begins to clear;][]{Owenetal2011}. They had the same range of $\delta$ and $R_\mathrm{CFE}$ values, with 10 radial bins but with a larger $\mathop{\mathrm{d}t}=0.175\,\mathrm{Myr}$, since the number of chondrule/dust populations that can be handled by the code is limited by the available memory. The $A$ values explored thus had to be lower (0.01--0.04) than in the steady-state runs to avoid over-processing chondrules in single CFEs.

\begin{figure*}\includegraphics[width=\textwidth]{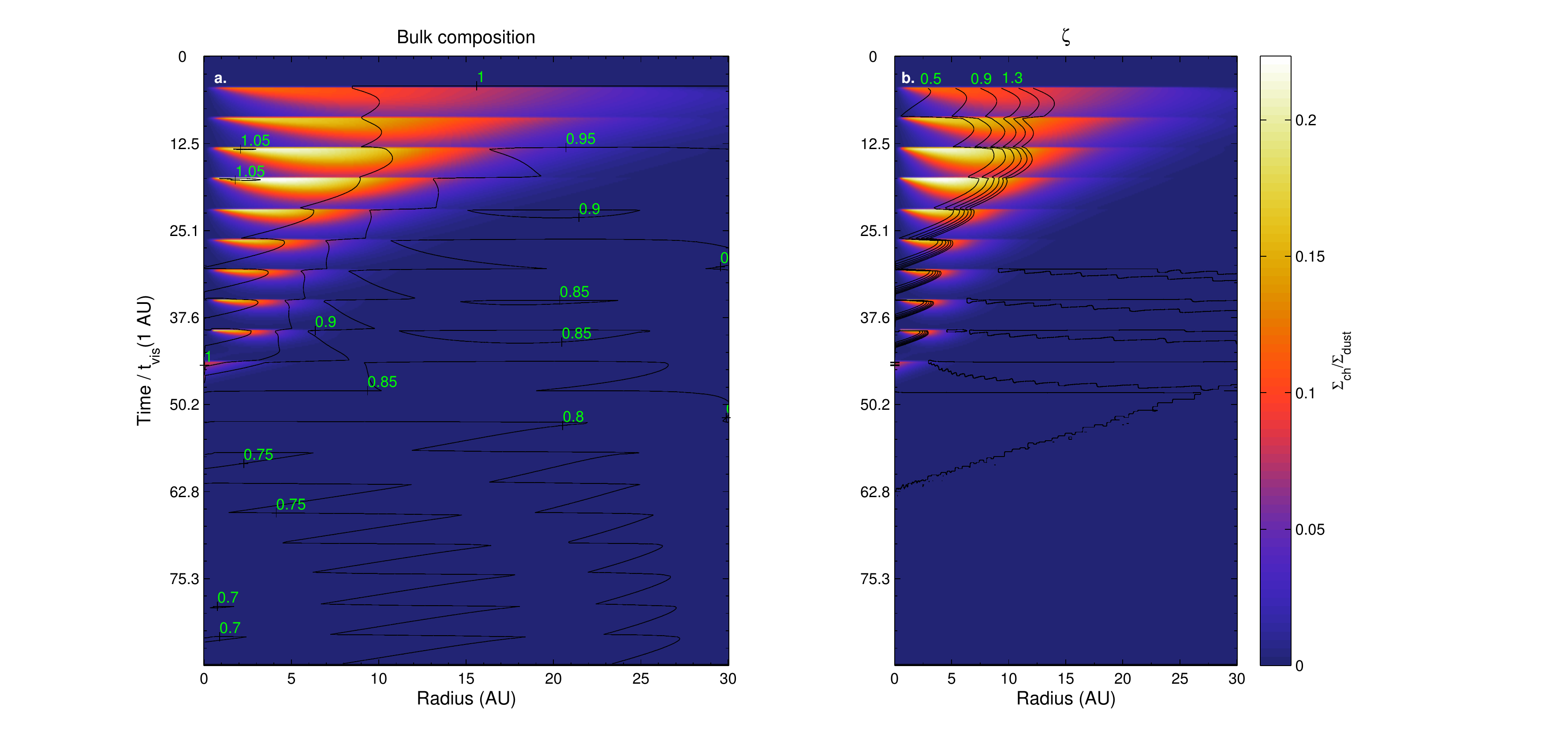}
		\caption{Space-time plot of the chondrule-to-dust surface density ratio, overplotted with contours of \textbf{a.} constant CI-normalized bulk composition and \textbf{b.} constant $\zeta$. This simulation had $\delta=0$, $R_\mathrm{CFE}=25\,\mathrm{AU}$, $A=0.04$, and the conservative prescription. Chondrules occupy a large portion of the inner disk, and the outer radius of the area they occupy decreases as time progresses. Contours with near-solar bulk composition and $\zeta$ within the complementary range robustly overlap with the high $\Sigma_{\mathrm{ch}}/\Sigma_{\mathrm{d}}$ region. After $\sim 2\,\mathrm{Myr}$, $\dot{M}$ is sufficiently small such that chondrules do not remain in the disk for very long.}
		\label{evolving-spacetime-example}
	\end{figure*}

	Figure \ref{evolving-spacetime-example} shows an example space-time diagram of the results of this simulation, using the same values of $\dot{M}$, $R_\mathrm{CFE}$, and $\delta$ as in Figure \ref{small-delta ex2}, and the largest value of $A$. Even though CFEs constantly occur, the area containing the majority of chondrule material continuously moves to smaller radii (along with contours of constant bulk composition and constant $\zeta$) as time progresses. This corresponds to the decreasing mass accretion rate onto the central star, explaining why exceedingly few chondrules are present at later times (after $\sim 2\,\mathrm{Myr}$, i.e. for $\dot{M}\lesssim 5\times 10^{-9}\:\mathrm{M_\odot/yr}$), in accordance with observations from the steady disk trials.

\begin{figure*}\includegraphics[width=\textwidth]{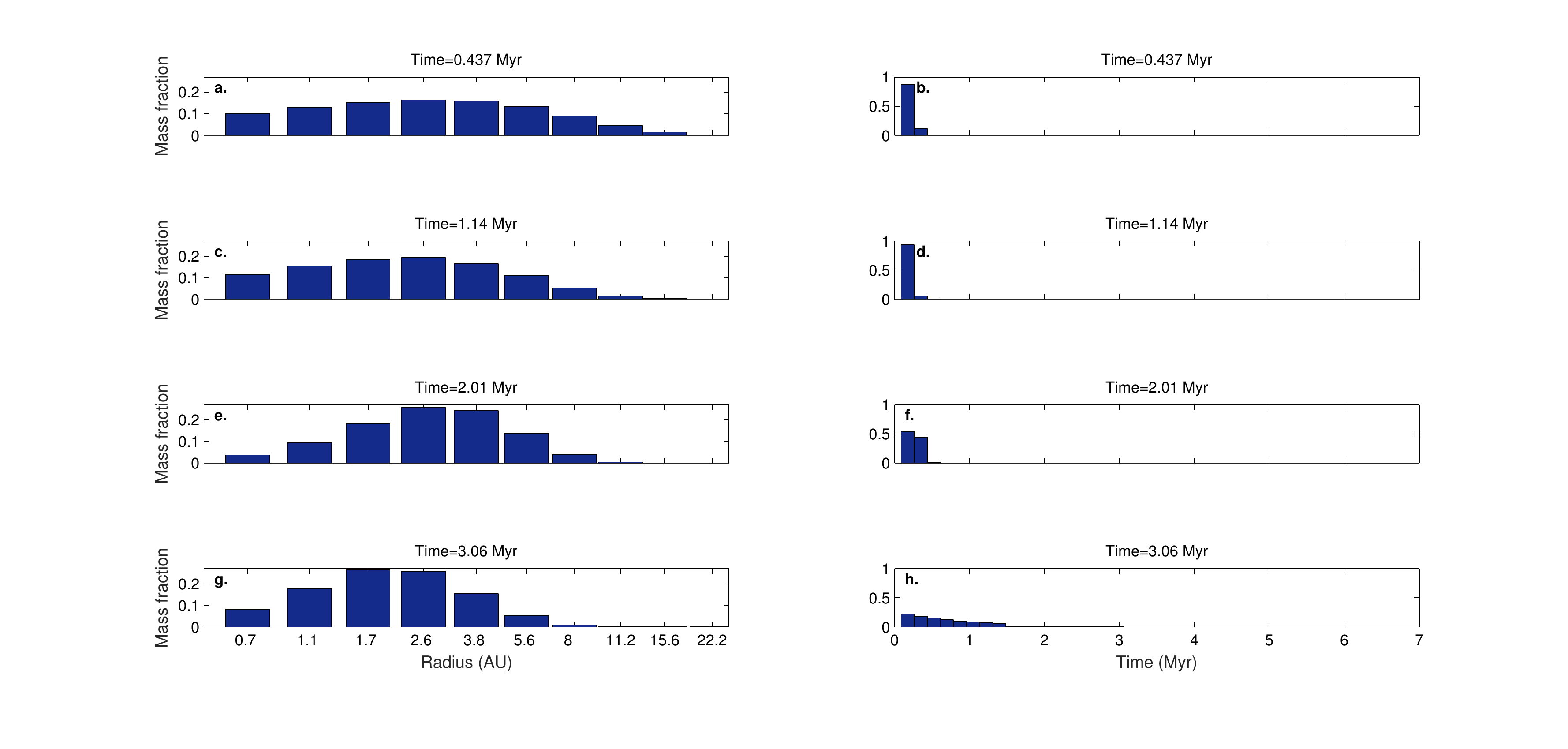}
		\caption{\textbf{Distributions of chondrule origins at} $R=3\,\mathrm{AU}$, in terms of mass fraction, after \textbf{a.} \& \textbf{b.} $t=0.437\,\mathrm{Myr}$ ($11.0 t_\mathrm{vis}(1\,\mathrm{AU})$), \textbf{c.} \& \textbf{d.} $t=1.14\,\mathrm{Myr}$ ($28.6 t_\mathrm{vis}(1\,\mathrm{AU})$), \textbf{e.} \& \textbf{f.} $t=2.01\,\mathrm{Myr}$ ($50.6 t_\mathrm{vis}(1\,\mathrm{AU})$), and \textbf{g.} \& \textbf{h.}  $t=3.06\,\mathrm{Myr}$ ($77.0 t_\mathrm{vis}(1\,\mathrm{AU})$), each of which are $0.05\,\mathrm{Myr}$ after a CFE. Origin radius is taken as the middle grid cell of each chondrule formation bin. Chondrules are seen to have mostly been formed near their current radius, with a greater emphasis towards smaller radii at larger times. There are chondrules present at $3\,\mathrm{AU}$ that have been formed at all times after $1.5\,\mathrm{Myr}$, with an emphasis towards recent formation.}
		\label{histograms-evolving}
	\end{figure*}

\begin{figure*}\includegraphics[width=\textwidth]{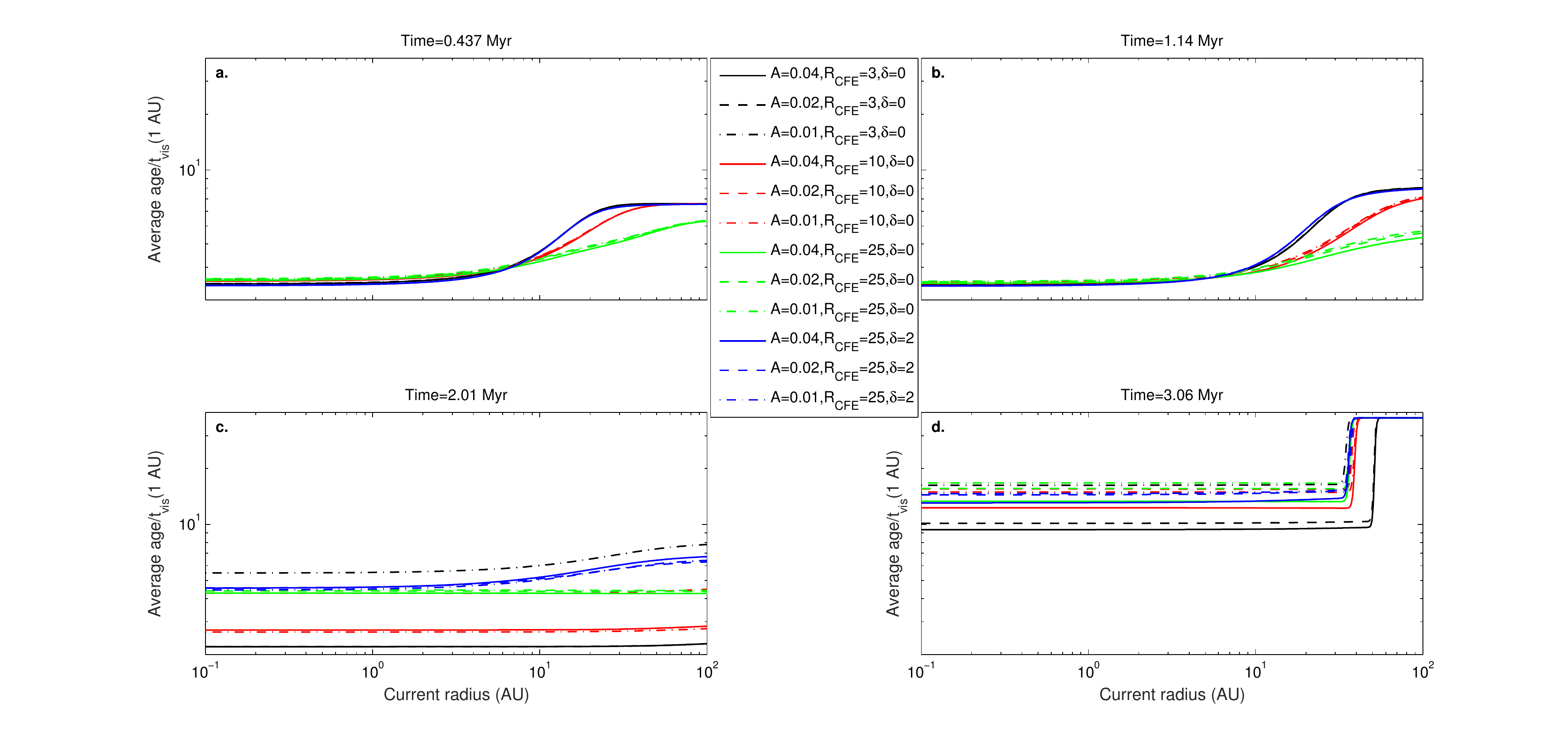}
		\caption{Average chondrule age normalized to $t_{\mathrm{vis}}(1\,\mathrm{AU})$ plotted versus $R$. Each plot occurs midway between the two closest CFEs, with \textbf{a.} $t=0.437\,\mathrm{Myr}$ ($11.0 t_\mathrm{vis}(1\,\mathrm{AU})$), \textbf{b.} $t=1.14\,\mathrm{Myr}$ ($28.6 t_\mathrm{vis}(1\,\mathrm{AU})$), \textbf{c.} $t=2.01\,\mathrm{Myr}$ ($50.6 t_\mathrm{vis}(1\,\mathrm{AU})$), and \textbf{d.} $t=3.06\,\mathrm{Myr}$ ($77.0 t_\mathrm{vis}(1\,\mathrm{AU})$). 
		Average age is smallest and constant at small $R$, before smoothly increasing; the onset of this increase correlates with the age of the disk. The value of $A$ is seen to be negligibly affect age at most times. In \textbf{a.} and \textbf{b.}, an increasing value of $R_{CFE}$ leads to a more gradual increase in average age, and the $\delta=2$ case coincides with the $R_{CFE}=3\,\mathrm{AU}$ distribution. In \textbf{d.}, it is observed that few young chondrules are seen at radii beyond $30$-$50\,\mathrm{AU}$.}
		\label{age-vs-R-evolving}
	\end{figure*}

\begin{figure*}\includegraphics[width=\textwidth]{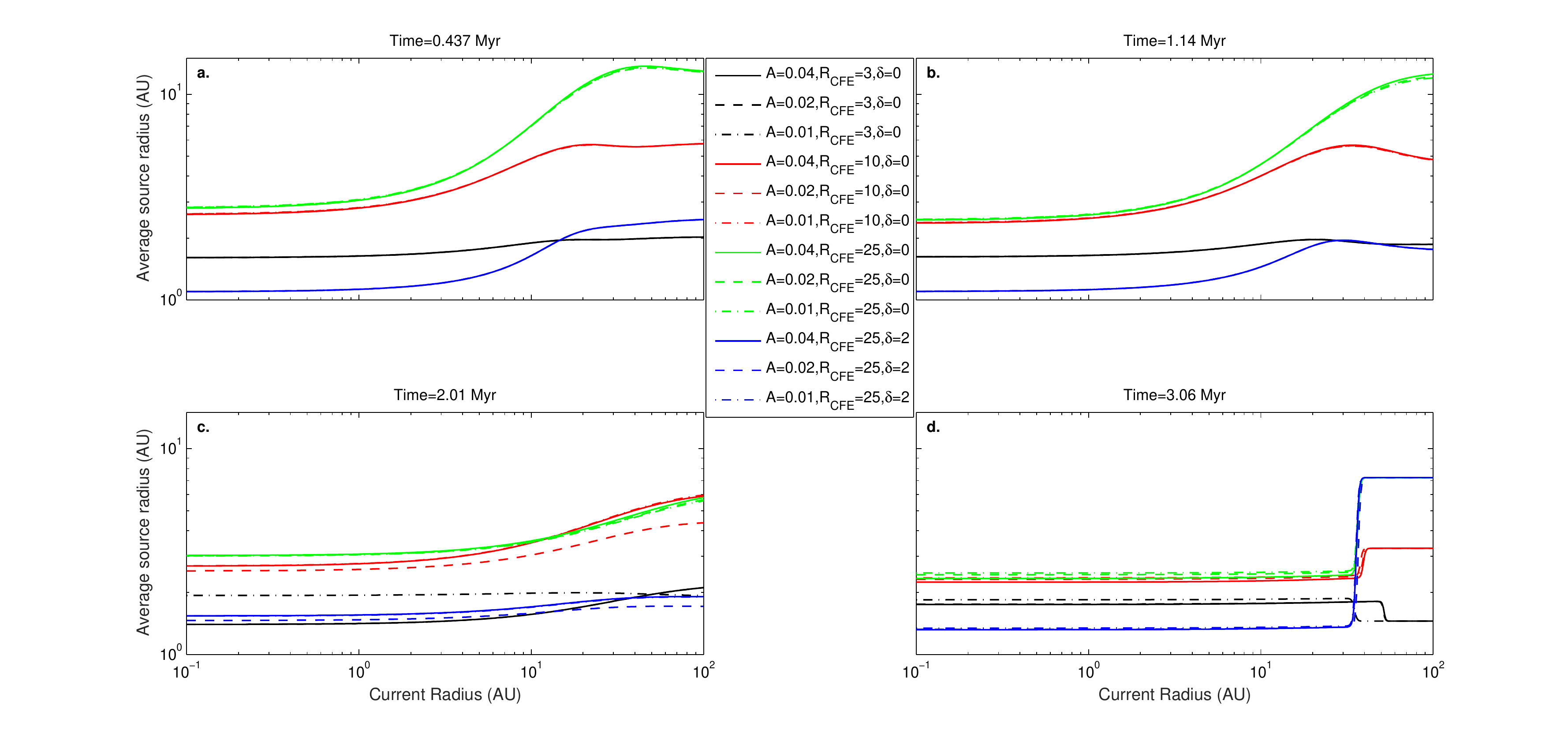}
		\caption{Average chondrule source radius plotted versus $R$ at \textbf{a.} $t=0.437\,\mathrm{Myr}$ ($11.0 t_\mathrm{vis}(1\,\mathrm{AU})$), \textbf{b.} $t=1.14\,\mathrm{Myr}$ ($28.6 t_\mathrm{vis}(1\,\mathrm{AU})$), \textbf{c.} $t=2.01\,\mathrm{Myr}$ ($50.6 t_\mathrm{vis}(1\,\mathrm{AU})$), and \textbf{d.} $t=3.06\,\mathrm{Myr}$ ($77.0 t_\mathrm{vis}(1\,\mathrm{AU})$). For all times, average source radius is approximately the same and constant over small radii. For the first three times, average source radius increases smoothly across the CFR, with the increase beginning at a larger radius for later times. In \textbf{d.}, source radius is constant up to $R=30$-$40\,\mathrm{AU}$, where it sharply increases to some average CFR radius. Increasing $R_{CFE}$ modestly increases average source radius at all times, and the value of $A$ only affects source radius at later times when the disk is depleted of chondrules. 
		}
		\label{space-vs-R-evolving}
	\end{figure*}

	Figure \ref{histograms-evolving} displays the histograms of source radii and epochs at 3 AU at four different times. The histograms are fairly comparable, at first glance, to their steady-state counterparts (see in particular the t=1.14 Myr panel, corresponding to $\dot{M}=10^{-8}\:\mathrm{M_\odot/yr}$, to be compared with Figure \ref{steady-histograms}). There is relatively little evolution in the first 2 Myr, in the direction of narrowing in agreement with trends for decreasing $\dot{M}$ discussed in the previous subsection. This can be seen more generally in Figures \ref{age-vs-R-evolving} and \ref{space-vs-R-evolving}, which display the average ages and source radii as functions of heliocentric distance for the different simulations (similar to Figures \ref{comparing-A-combined} and \ref{combined-vs-S-bigger} in the steady-state case) at the same four times as Figure \ref{histograms-evolving}. We note a downturn in average source radius further outward, which may reflect a significant contribution of chondrules produced close to the Sun transported there by the original expansion of the disk, an effect seen by \citet{YangCiesla2012} in the context of refractory inclusions. This may explain the somewhat shorter source radii and older ages overall relative to the steady-state runs. This contribution of early-transported chondrules becomes dominant at 3 Myr when newly produced chondrules are rapidly lost, as may be seen in the final panel of Figure \ref{age-vs-R-evolving}, although all chondrules in general are scarce at that point. There is also no dip in age in Figure \ref{age-vs-R-evolving} unlike the steady-state runs. This is presumably partly an artifact of a larger $\mathop{\mathrm{d}t}$, but may also relate to the inflow of the early-transported chondrules from further out.

	\section{Cosmochemical implications}
	\label{section-discussions}
	
	\subsection{Matrix-chondrule relationships}
	
	One first conclusion of the above study (in particular Section \ref{Dust top hat model}) is that a complementary relationship between cogenetic chondrules and dust can be preserved for long timescales provided that the decoupling between chondrules and gas is limited, that is $S\ll 1$. This should hold in the inner solar system for relatively high mass accretion rates ($\dot{M}\gtrsim 10^{-7}\:\rm M_\odot/yr$), viz., early in the disk evolution as our continuous chondrule formation simulations (Section \ref{section-continuous-CFE}) verify. We also note that low mass accretion rates $\dot{M}\lesssim 10^{-9}\:\rm M_\odot/yr$ would likely not allow chondrules to be retained efficiently in the disk because of radial drift, depending on their production rate. Thus evidence of matrix-chondrule complementarity in carbonaceous chondrites \citep[e.g.][]{HezelPalme2010,Palmeetal2015} is not inconsistent with significant transport between \textit{chondrule formation} and \textit{chondrite accretion} \citep{Jacquetetal2012S}. There is further no inconsistency with evidence for the presence of a CI chondritic component in carbonaceous chondrites \citep{Anders1964,Zandaetal2006,Zandaetal2012}, as exemplified by our modelling of primordial CI dust that has been mixed with processed matrix components.  This does not however prejudge the possibility that the empirical evidence of matrix-chondrule complementarity \textbf{found in carbonaceous chondrites}, e.g. as to the Mg/Si ratio, is compromised by parent body alteration or instrumental biases \citep{Zandaetal2012}.
	
	Given that higher levels of gas-solid decoupling (higher $S$) do lead to chondrule/dust fractionation, it is then tempting to associate the nonsolar compositions of the noncarbonaceous chondrites with this effect, which was one of the suggestions by \citet{Jacquetetal2012S}. This would, however, be difficult for the Mg/Si ratio. Indeed, non-carbonaceous chondrites display a subsolar Mg/Si ratio, while the higher-Mg/Si chondrite components are the chondrules themselves. Our simulations suggest that the innermost regions of the disk would be enriched in chondrules and \textbf{give rise to chondrites with} an \textit{enhanced} Mg/Si ratio, unlike observations. True, a correspondingly low-Mg/Si region may appear further out, but the effect would be comparatively limited, and would predict that non-carbonaceous chondrites are depleted in chondrules relative to carbonaceous chondrites, the contrary of observations. The transport of chondrules by itself is thus unlikely to have generated the difference between carbonaceous and non-carbonaceous chondrites, although it could explain part of the diversity internal to non-carbonaceous chondrites, for example regarding metal/silicate fractionation \citep{Zandaetal2006,Vernazzaetal2014}. A remaining possibility, independent of chondrule formation, to be investigated in the future, is the loss of amoeboid olivine aggregates \citep[a class of refractory inclusions;][]{LarimerWasson1988refractory,Jacquet2014review}.
	
	We have ignored settling effects in case accretion takes place preferentially at the midplane. While this would incur little change for $S\ll 1$ (and thus not affect the complementarity argument above), this would increase the chondrule/dust ratio in chondrites above the surface density ratio \citep{Jacquetetal2012S}. But, if accretion is significant, the disk, becoming preferentially depleted in chondrules, would become richer in dust, which might invert the trend. Another possibility is that dust accreted on chondrules as rims before final agglomeration \citep{Metzleretal1992,Ormeletal2008}, so that representativity of chondrites vis-\`{a}-vis the whole nebular reservoir would be more faithful.  
	
	\subsection{Space-time distribution of chondrules}
	
	Taken at face value, our simulations (with $\alpha=0.0025$) indicate that the source regions of chondrules of a given chondrite would be dominantly in the neighbourhood of \textbf{the chondrite's} accretion region, with a radius standard deviation equal to a significant fraction of the overall width of the CFR in the inner disk and smaller further out. One could accommodate a few distinct contemporaneous chondrite groups along the CFR, but certainly not $>14$ as observed \citep{Jones2012}. But since the typical age of chondrules would be of order $t_{\rm vis}$ (here $\ll$ 1 Myr), there would be ample room for some of the chondrite diversity to reflect diversity in \textbf{chondrite} accretion time (in addition to heliocentric distance). We further note that the narrow age range would be dictated by disk dynamics and, if confirmed empirically, would thus not imply immediate chondrite accretion following chondrule formation (contra \citet{AlexanderEbel2012}).
	
	Yet the radiochronological data suggest a wider age range (a few Myr) for chondrules in individual chondrites \citep{Villeneuveetal2009,Connellyetal2012}, although the possibility of secondary disturbances is not yet ruled out \citep{AlexanderEbel2012}. This might simply indicate that the $\alpha$ value used (0.0025) is too high and that the viscous timescale is more comparable to a few Myr ($\alpha\lesssim 10^{-4}$), as might be expected in a dead zone currently believed to encompass a large fraction of the planet-forming region of protoplanetary disks \citep{Gammie1996}. (Recall that changing the $\alpha$ for a given mass accretion rate merely amounts to rescaling the time so our results expressed in terms of $t_{\rm vis}$ remain valid.) It could well be that a steady state in terms of chondrule relative radius-age distribution was not yet reached, so that more spatial diversity (due to smaller diffusion length) can be envisioned. This would be compatible with a temporal evolution as well \citep{Jacquetetal2012S}. The observed preservation of refractory inclusions, which formed early in the solar system \citep[perhaps within the first $0.1\,\mathrm{Myr}$; e.g.][]{Bizzarroetal2004,Amelinetal2010}, in large abundances in CCs may be a further argument in favour of low $\alpha$ values \citep{Jacquetetal2011a}. Still, it should be noted that simulations of refractory inclusion transport by \citet{YangCiesla2012} achieve suitable levels of preservation assuming a moderately high $\alpha=10^{-3}$, \textit{provided} the disk was initially very compact ($R_1\lesssim 10\:\rm AU$) so that its viscous expansion sent many refractory inclusions far from the Sun (10--100 AU), hence a longer drift timescale afterward. In the current understanding of disk magnetohydrodynamics, a dead zone would nonetheless have rapidly emerged anyway, strengthening the reported retention.
	
	A further noteworthy constraint is provided by analyses of dust returned from comet Wild 2 by the Stardust mission. Indeed, while a carbonaceous chondrite-like component is certainly present, oxygen isotopes \citep{Nakashimaetal2012} and chemical compositions \citep{Franketal2014} of olivine grains show that the contribution of most carbonaceous chondrite groups is limited, and that CR chondrites as well as non-carbonaceous groups \citep[the former showing some characteristics transitional with the latter;][]{JacquetRobert2013} may be significant sources. This is at variance with the idea that non-carbonaceous and carbonaceous chondrites only differ by their spatial origins, with carbonaceous chondrites being furthest, since the spatial source distribution should converge in the outer solar system (see Section \ref{Steady disk simulations}), so that in terms of CFR-originated components, comets should be dominated by carbonaceous chondrite ones. The Stardust mission may thus offer further evidence for a temporal evolution, with possibly Wild 2 being the result of a relatively late accretion \citep{Nakashimaetal2012}, if the suggestion by \citet{Jacquetetal2012S} that noncarbonaceous chondrites \textbf{accreted} later than their carbonaceous counterparts holds. This may also explain the lack of evidence of initial live $^{26}$Al for several Wild 2 particles \citep[e.g.][]{Oglioreetal2012}. Carbonaceous chondritic material may have survived longer at greater distance, e.g. in the wake of an initial expansion of the disk \citep{YangCiesla2012}.

	\section{Conclusions}
	\label{section-conclusions}
We have performed a  numerical investigation of chondrule transport in the young Solar System. Our results speak to the nature of the chondrule-matrix relationship; spatial and temporal constraints on the chondrule formation process; and the success of deploying astrophysical numerical methods for cosmochemical applications. The main results are as follows:
	
\begin{enumerate}

\item We have defined a ``complementarity parameter'' ($\zeta$) as a metric for comparing the chondrule-matrix relationships in our astrophysical models to those found in lab-analysed chondrites. It evaluates how close the chondrule and matrix component are to being complementary, while being independent of how close the individual components are to having solar abundances.  

\item Our simulations showed that the gas-solid decoupling parameter $S$ (the ratio of the diffusion to drag timescales)  is predominantly responsible for these relationships, with sufficiently low values of $S$ (the diffusion dominated regime) allowing for the time between chondrule production and chondrite accretion to be longer than the disk's viscous timescale. Older disks, with lower mass accretion rates, are heavily depleted of chondrules relative to their younger counterparts. Thus, we constrain the chondrule formation events to have occurred early, in a disk with $\dot{M}\gtrsim 10^{-9}$~M$_\odot$~yr$^{-1}$ for nominal parameters.
	
\item At a given radius and time in the planetary disk, chondrule origin varies as a function of space and time, each of which are affected by disk parameters as well as the rate of chondrule formation. Again, mass accretion rate, and thus disk age, is seen to be relevant to chondrule distributions. 

\item The location of the $S=1$ line limits the outward diffusion of chondrule matter. The distribution and extent of chondrule formation in space significantly impacts spatiotemporal diversity, because most chondrules accreted in the inner disk formed locally. 

\item The spatial diversity in our simulations is insufficient to explain the number of distinct chondrite classes that exist, but we argue that temporal diversity can, and likely does, account for this discrepancy. More radiochronological data establishing temporal bounds on chondrite classes in terms of either chondrule age or chondrite accretion time are obviously needed.
\end{enumerate}	

This study shows the viability of our numerical methods in investigating chondrule origins.  The agreement between our simulations for a single CFE, continuous CFEs with a static disk, and continuous CFEs with an evolving disk convey the power of this technique for exploring regimes unattainable through pure analytics or cosmochemical speculation, and we note that such simulations are {\it not} computationally expensive. Furthermore, our simulations make predictions of their own, which can be candidly evaluated in future studies.  This successful marriage of a cosmochemical quandary to an astrophysical method shows the strength of this relationship, and we ardently promote the exploitation of similar relationships in future endeavours.
	
	\section*{Acknowledgments}
	We thank the anonymous referee for insightful comments that improved this paper.
 AZG acknowledges support of an NSERC summer research grant in the framework of the Summer Undergraduate Research Program held at CITA in 2014. JEO acknowledges support by NASA through Hubble
Fellowship grant HST-HF2-51346.001-A awarded by the
Space Telescope Science Institute, which is operated by
the  Association  of  Universities  for  Research  in  Astronomy, Inc., for NASA, under contract NAS 5-26555.	The calculations were performed on the Sunnyvale cluster at CITA,
	which is funded by the Canada Foundation for Innovation.
	
	\bibliographystyle{mn2e}
	\bibliography{bibliography}
	
	\begin{appendix}

		\section{Calculation of $t_{\rm half}$ in the regime $1< S< R_\mathrm{centre}/L$}
		\label{Jacquet regime}
		In this regime (as can be verified {\it a posteriori}), the initial widths of the chondrule and dust populations can be neglected and the variation of heliocentric distance can be ignored. After a time $t$, the surface density profiles of chondrules and dust can be approximated by a Gaussian distribution:
		\begin{equation}
		\Sigma_{\mathrm{ch,d}}(R)=\frac{M_\mathrm{ch,d}}{\sqrt{2\pi}\sigma}\exp\left(-\frac{ \left(R-R_{\mathrm{centre}_\mathrm{ch,d}}\right)^2}{2\sigma^2}\right),
		\end{equation}
		where $\sigma=\sqrt{2Dt}$ for both populations, and $M_{\rm ch}$ and $M_{\rm d}$ are the masses of the chondrule and dust populations, respectively. The complementarity parameter then reduces to:
		\begin{equation}
		\zeta=\frac{\Sigma_{\mathrm{ch}}/\Sigma_{\rm d}}{M_{\rm ch}/M_{\rm d}}=\exp\left(\frac{-2Rd+d\left(R_{\mathrm{centre}_\mathrm{ch}}+R_{\mathrm{centre}_\mathrm{d}}\right)}{2\sigma^2}\right),
		\end{equation}
		with
		\begin{equation}
		\begin{split}
		d\equiv R_{\mathrm{centre}_\mathrm{d}}-R_{\mathrm{centre}_\mathrm{ch}}=\left(v_{\mathrm{drift}_\mathrm{d}}-v_{\mathrm{drift}_\mathrm{ch}}\right) t \\
		\approx \lvert v_{\mathrm{drift}_\mathrm{ch}}\rvert t.
		\end{split}
		\end{equation}
		from which one can obtain the abscissas corresponding to $\zeta=a$ and $\zeta=b$. Integrating then the chondrule surface density in between yields
		
		\begin{equation}
		w_{a \leq \zeta \leq b}=\frac{\operatorname{erf}\left( \frac{\ln{b}}{\sqrt{2}}\frac{\sigma}{d}-\frac{1}{2\sqrt{2}}\frac{d}{\sigma}\right)-\operatorname{erf}\left( \frac{\ln{a}}{\sqrt{2}}\frac{\sigma}{d}-\frac{1}{2\sqrt{2}}\frac{d}{\sigma}\right)}{2},
		\end{equation}
		giving $\frac{\sigma}{d}\approx 2.297$ for $\left(w,a,b\right)=\left(0.5,0.7,1.3\right)$. Plugging $t_\mathrm{half}$ into the definitions of $\sigma$ and $d$ and substituting for $v_{\mathrm{drift}_\mathrm{ch}}$ gives
		\begin{equation}
		t_\mathrm{half}=\frac{2D}{2.297^2}\left(\frac{\rho}{\tau\cdot \partial P/\partial R}\right)^2.
		\end{equation}

	\end{appendix}
	
	\label{lastpage}
\end{document}